\documentclass[]{pasj02} 
\usepackage[switch,mathlines]{lineno} 
\usepackage{natbib} 
\usepackage{url}

\nolinenumbers

\jyear{2024}
\Received{}
\Accepted{}

\usepackage{amsmath}

\usepackage{ulem}
\usepackage{soul}
\setul{}{1pt}
\definecolor{kocolor}{cmyk}{1,0,0,0.3}
\definecolor{yycolor}{cmyk}{0,0.6,0.8,0.1}
\definecolor{tkscolor}{cmyk}{0,1,0,0.2}

\def\mbf#1{\mbox{\boldmath ${#1}$}}

\graphicspath{{./}{figures/}{../work/}}

\begin{document}

\title{Red Giant Winds Driven by Alfv\'en Waves with Magnetic Diffusion}

%
%

\author{
 Takeru K. \textsc{Suzuki},\altaffilmark{1}\altemailmark\orcid{0000-0001-9734-9601} \email{stakeru@ea.c.u-tokyo.ac.jp} 
 Keiichi \textsc{Ohnaka},\altaffilmark{2}\orcid{0000-0000-0000-0000}
 and
 Yuki \textsc{Yasuda},\altaffilmark{3}\altemailmark\orcid{0000-0003-1093-0052} 
}
\altaffiltext{1}{School of Arts \& Sciences, The University of Tokyo,  3-8-1, Komaba, Meguro, Tokyo 153-8902, Japan; Department of Astronomy, The University of Tokyo, 7-3-1,
  Hongo, Bunkyo, Tokyo, 113-0033, Japan; Komaba Institute for Science, The University of Tokyo, 3-8-1 Komaba, Meguro, Tokyo 153-8902, Japan}
\altaffiltext{2}{Instituto de Astrof\'isica, Departamento de F\'isica y Astronom\'ia, Facultad de Ciencias Exactas, Universidad Andr\'es Bello, Fern\'andez Concha 700, Las Condes, Santiago, Chile}
\altaffiltext{3}{Division of Physics, Faculty of Science, Kita 10 Nishi 8, Kita-ku, Hokkaido University, Sapporo 060-0810, Japan}


\maketitle

\begin{abstract}
  We investigate the driving mechanism of Alfv\'en wave-driven stellar winds from red giant stars, Arcturus ($\alpha$ Boo; K1.5 III) and Aldebaran ($\alpha$ Tau; K5 III), with nonideal magnetohydrodynamical (MHD) simulations in one-dimensional super-radially open flux tubes. Since the atmosphere is not fully ionized, upward propagating Alfv\'enic waves excited by surface convection are affected by ambipolar diffusion. Our fiducial run with the nonideal MHD effect for $\alpha$ Boo gives a time-averaged mass-loss rate, $\langle\dot{M}\rangle=3.3\times 10^{-11}M_{\odot}$yr$^{-1}$, which is more than one order of magnitude reduced from the result in the ideal MHD run and nicely explains the observational value. Magnetized hot bubbles with temperature, $T\gtrsim 10^6$ K are occasionally present simultaneously with cool gas with $T\sim$ a few thousands K in the atmosphere because of the thermal instability triggered by radiative cooling; there coexist fully ionized plasma emitting soft X-rays and molecules absorbing/emitting infrared radiations.
  The inhomogeneity in the atmosphere also causes large temporal variations in the mass-loss rate within an individual magnetic flux tube. 
  We also study the effect of magnetic field strength and metallicity, and find that the wind density, and accordingly the mass-loss rate, positively and sensitively depends on both of them through the ambipolar diffusion of Alfv\'enic waves.
  The nonideal MHD simulation for $\alpha$ Tau, which is slightly more evolved than $\alpha$ Boo and has weaker magnetic field, results in weaker wind with $\langle\dot{M}\rangle=1.5\times 10^{-12}M_{\odot}$yr$^{-1}$ with the atmospheric temperature $\lesssim 10^5$ K throughout the simulation time. 
  However, given the observations implying the presence of locally strong magnetic fields on the surface of $\alpha$~Tau, we also conduct a simulation with a field strength twice as strong. This results in $\langle\dot{M}\rangle=2.0\times 10^{-11}M_{\odot}$yr$^{-1}$ -- comparable to the observed value -- with transient magnetized hot bubbles.
\end{abstract}



\section{Introduction} \label{sec:intro}
Stellar winds are blowing out from stars and their physical properties show large diversity depending on the stellar type. The solar wind is driven by magnetohydrodynamical (MHD) processes; Alfv\'en(ic) waves excited by surface convection \citep[e.g.,][]{Suzuki2005,Cranmer2005,Matsumoto2012,Matsumoto2021} and magnetic activities \citep[][]{Axford1992,Zank2020,Iijima2023} are a promising source to accelerate the wind.
As the Sun and solar-type stars evolve, they expand,  and the mass-loss rate due to stellar winds drastically increases from $\dot{M}\sim 10^{-14} - 10^{-12}M_\odot$yr$^{-1}$ during the main sequence (MS) phase 
\citep{Wood2005,Cranmer2011} to  
$\dot{M}\sim 10^{-4}M_\odot$yr$^{-1}$ in the final stage on the asymptotic giant branch (AGB) phase \citep{Renzini1981,Engels1983,Baud1983,vanLoon2005}. 
It is argued that the extremely large mass-loss rate of AGB stars is induced by a combination of the levitation of atmospheric gas due to stellar pulsation and/or convection and the radiation pressure on dust grains \citep[][]{Bowen1988,Hoefner2018,Freytag2023}. 

Between the MS and AGB phases, stars are in the red giant branch (RGB) phase before the helium is ignited in the stellar core. 
With the evolution in the RGB phase, the properties of the atmosphere and wind drastically change. Stars in the blueward side of ``the dividing line'' \citep{Linsky1979, Ayres1981} tend to have stable hot corona where the wind properties such as terminal velocity are difficult to be determined observationally. On the other hand, late-type giants that are redward of the dividing line 
mostly do not show clear signatures of a stable, hot corona. 
{However, non-coronal giants still possess a moderately warm atmosphere. For example, \citet{McMurry1999} introduced a semiempirical hydrostatic model for the atmosphere of Aldebaran ($\alpha$ Tau; K5 III) that explains UV lines observed with HST. A remarkable ingredient in the model 
is to include a transition region with $T\approx 10^5$ K at $r\approx 1.2 R_{\star}$ in addition to the chromosphere \citep[c.f.][for a traditional model]{Kelch1978}. }

The wind properties of nearby non-coronal giants have been estimated based on analyses of chromospheric and transition region lines \citep[e.g.,][]{Ayres1982, Robinson1998, Wood2016, Harper2022}; typical non-coronal K giants Arcturus ($\alpha$ Boo; K1.5 III) and Aldebaran ($\alpha$ Tau; K5 III) exhibit moderately large mass-loss rates, $\dot{M}\sim 10^{-11}M_\odot$yr$^{-1}$ \citep{Robinson1998,Harper2022}.
However, the driving mechanism of the stellar wind is still uncertain. While 
{dust-driven winds probably do not work} as the temperature in the atmosphere is still too high to form dust grains, the MHD mechanism is 
regarded to be a viable candidate as non-coronal K giants possess magnetic fields as a result of the dynamo mechanism in the surface convective layer \citep{Auriere2015,Schroeder2018}.
{\citet{Wedemeyer2017}, who only found a cool atmosphere with $T\lesssim 5000$ K in their three-dimensional radiation hydrodynamical simulations without magnetic fields, also argue the importance of magnetic heating in the formation of the warm chromosphere on non-coronal giants.}

An important point regarding MHD processes in the atmosphere of non-coronal giants is that the gas is generally only weakly ionized. Therefore, the magnetic field is not tightly coupled to the gas and could diffuse out \citep{Holzer1983}. \citet{Airapetian2010} performed multi-dimensional MHD simulations with resistivity and reported that broadband Alfv\'en waves explain general properties of the stellar wind from RGB stars.
\citet{Yasuda2019} explicitly took into account the ambipolar diffusion between the neutral component and the magnetic field coupled to the charged component in addition to the resistivity, and studied the evolution of the stellar wind from the RGB to the AGB with one-dimensional nonideal MHD simulations.
While the current paper basically follows \citet{Yasuda2019}, we focus on the effect of magnetic diffusion on stellar winds from non-coronal K giants and examine the detailed physical properties including temporal variability and the dependence on magnetic field and metallicity.  

The paper is organized as follows. In Section \ref{sec:method}, we present the setup for the nonideal MHD simulations. Section \ref{sec:massloss} presents the dependence of mass-loss rates on stellar evolution. In Section \ref{sec:Boo} we extensively discuss the results for $\alpha$ Boo including the dependence on magnetic field strength and metallicity. In Section \ref{sec:Tau}, we describe the results for $\alpha$ Tau. Section \ref{sec:sum} summarizes the paper.

\section{Models and Simulation Methods}
\label{sec:method}

\begin{table}
  \tbl{Stellar properties adopted in this work.}{
    \begin{tabular}{c|ccc}
      \hline
      \hline
      Property & $\alpha$ Boo & $\alpha$ Tau & Note/Ref.\footnotemark[$*$] \\
      \hline
      $M_{\star} [M_{\odot}]$ & 1.0 & 1.3 & [1,2]\\
      $R_{\star} [R_{\odot}]$ & 25.4 & 45.2 & [2--4]\\
      $T_\mathrm{eff}$ [K]& 4300 & 3900 & [1,5--9]\\
      $L_{\star} [L_{\odot}]$ & 195 & 418 & from $R_{\star}$ \& $T_\mathrm{eff}$ \\
      $\log g$ & 1.6 & 1.2 &  from $M_{\star}$ \& $R_{\star}$ \\
      $\rho_{0}$[g cm$^{-3}$] & $9.4\times 10^{-9}$ & $6.0\times 10^{-9}$ & eq. (\ref{eq:rho0})\\
      $c_\mathrm{s,0}$ [km s$^{-1}$]& 5.4 & 5.1 & eq. (\ref{eq:p0})\\
      $\delta v_0$[km s$^{-1}$] & 2.50 & 2.56 & eq. (\ref{eq:dv0})\\      
      $H$ [km s$^{-1}$] & $6.8\times 10^4$ & $1.5\times 10^5$ & eq. (\ref{eq:h/R})\\
      $\omega_\mathrm{max}^{-1}$[min] & 150 & 340 & eq. (\ref{eq:wvP})\\
      $B_{r,0}$ [G] & 262 & 198 & eq. (\ref{eq:eqP})\\
      $f_0B_{r,0}$ [G] & 0.23 - 0.91 & 0.13 - 0.38 & [7,8] \\
      $h/R_{\star}$ & 0.55 & 0.66 & eq. (\ref{eq:h/R})\\
      $\lambda_{\perp 0}$ [km] & $1.0\times 10^6$ & $2.2\times 10^6$ & \\ 
      \hline
    \end{tabular}    
  }
  \label{tab:stprm}
  \begin{tabnote}
    \footnotemark[$*$]
    Refs: [1]: \citet{Heiter2015},  [2]:\citet{Abia2012}, [3]: \citet{Lacour2008}, [4]: \citet{vanLeeuwen2007}, [5]: \citet{Wood2016}, [6]: \citet{Peterson1993}, [7]: \citet{Sennhauser2011}, [8]: \citet{Auriere2015}, [9]:\citet{Taniguchi2018}.
  \end{tabnote}
\end{table}

\begin{table}
  \tbl{Elemental abundances adopted in this work. }{
    \begin{tabular}{c|c|cc|c}
      \hline
      \hline
      Element\footnotemark[$*$] & Sun & $\alpha$ Boo & Note/Ref.\footnotemark[$**$]  & $\alpha$ Tau\\
      \hline
      $A(\mathrm{He})$ & $8.4\times 10^{-2}$ & $8.3\times 10^{-2}$ & [1] & $8.4\times 10^{-2}$\\
      $A(\mathrm{C})$ & $2.9\times10^{-4}$ & $1.2\times 10^{-4}$ & [2] & $2.1\times10^{-4}$\\
      $A(\mathrm{O})$ & $4.9\times 10^{-4}$ & $5.0\times10^{-4}$ & [2] & $3.6\times 10^{-4}$\\  
      $A(\mathrm{Na})$ & $1.7\times10^{-6}$ & $5.0\times 10^{-7}$ & $0.3Z_{\odot}$ & $1.3\times10^{-6}$\\ 
      $A(\mathrm{Mg})$ & $3.5\times10^{-5}$ & $3.1\times10^{-5}$ & [3] & $2.6\times10^{-5}$\\
      $A(\mathrm{Al})$ & $2.7\times 10^{-6}$ & $8.1\times 10^{-7}$ & $0.3Z_{\odot}$ & $2.0\times 10^{-6}$ \\
      $A(\mathrm{Si})$ & $3.2\times 10^{-5}$ & $1.7\times 10^{-5}$ & [3] & $2.4\times 10^{-5}$ \\
      $A(\mathrm{S})$  & $1.3\times 10^{-5}$ & $4.0\times 10^{-6}$ & $0.3Z_{\odot}$ & $9.6\times 10^{-6}$ \\
      $A(\mathrm{K})$ & $1.2\times 10^{-7}$ & $3.5\times 10^{-8}$ & $0.3Z_{\odot}$ & $8.9\times 10^{-8}$\\
      $A(\mathrm{Ca})$ & $2.0\times 10^{-6}$ & $6.0\times 10^{-7}$ & $0.3Z_{\odot}$ & $1.5\times 10^{-6}$ \\
      $A(\mathrm{Cr})$ & $4.2\times 10^{-7}$ & $1.3\times 10^{-7}$ & $0.3Z_{\odot}$& $3.1\times 10^{-7}$\\
      $A(\mathrm{Fe})$ & $2.9\times 10^{-5}$ & $8.5\times 10^{-6}$ & [3] & $2.1\times 10^{-5}$\\
      \hline
  \end{tabular}}
  \label{tab:metal}
  \begin{tabnote}
    \footnotemark[$*$]
    $A(\mathrm{X})$ represents the abundance of the element X relative to H in the number density.\\
    \footnotemark[$**$]
    Refs: [1]: \citet{Drake1985}, [2]: \citet{Ryde2010}, [3]:\citet{Jofre2015}. The solar abundances of the heavy elements are adopted from \citet{Asplund2021}, while the solar He abundance is from \citet{Asplund2009}. $A(X)$ of $\alpha$ Tau is scaled to the solar values via $0.74Z_{\odot}$, corresponding to [Fe/H]$=-0.13$ as determined by \citet{Abia2012}.
  \end{tabnote}
\end{table}

Our numerical treatment is an extension of \citet[][M24 hereafter]{Matsuoka2024} that studied Alfv\'en wave-driven solar wind with magnetic diffusion.
We replace the Sun with a moderately evolved RGB star and investigate physical properties of the stellar wind.  For our standard case, we employ Arcturus ($\alpha$ Boo; K1.5 III) as a typical non-coronal K giant. The basic stellar parameters are summarized in Table \ref{tab:stprm}; the stellar mass, $M_{\star}$ and radius, $R_{\star}$, are adopted from the compilation by \citet{Harper2022}. For the effective temperature, $T_\mathrm{eff}$, we adopt a 'rounded' value averaged from multiple literatures. The luminosity, $L_{\star}=4\pi R_{\star}^2 \sigma T_\mathrm{eff}^2$,  and surface gravity, $g=GM_{\star}/R_{\star}^2$, are calculated from these stellar parameters, where $\sigma$ and $G$ are the Stefan-Boltzmann constant and the gravitational constant, respectively. 
We also simulate a slightly more evolved RGB star, Aldebaran ($\alpha$ Tau; K5 III), as presented in Table \ref{tab:stprm}, to examine the dependence of detailed wind properties on stellar evolution. 

\subsection{Elemental Abundances}
\label{sec:abnd}
The abundances of heavy elements are a key factor in determining the properties of winds from RGB stars because, in the weakly ionized atmosphere with $T<10^4$K, elements with low first ionization potential (FIP) are the main suppliers of electrons, which control magnetic diffusion (Section \ref{sec:Bdif}).
We consider the ionization of low-FIP elements, C, O, Na, Mg, Al, Si, S, K, Ca, Cr, and Fe, in addition to H and He.

Arcturus ($\alpha$ Boo) is moderately metal-poor with 
[Fe/H]$=-0.52$ \citep{Jofre2014}, which corresponds to the metallicity, $Z=0.3Z_{\odot}$, where $Z_\odot$ stands for the solar metallicity. The abundances of He, C, O, Mg, Si, and Fe are adopted from the compiled data by \citet{Harper2022}, where the original literatures are listed in Table \ref{tab:metal}. We scale the abundances of the other heavy elements from the solar values \citep{Asplund2021}, using $Z=0.3Z_{\odot}$.
The metallicity of Aldebaran ($\alpha$ Tau) is close to the solar value, [Fe/H]$=-0.13$ \citep{Abia2012}.  We scale the abundances of the heavy elements 
with $Z=0.74 Z_{\odot}$.

\subsection{Photosphere: Inner Boundary}
\label{sec:inbd}
The inner boundary of the numerical domain is set at the location where $T=T_\mathrm{eff}$.
The temperature is set to $T_\mathrm{eff}$ and the density, $\rho_0$, is scaled via
\begin{equation}
  \rho_0 \propto g^{0.6}T_\mathrm{eff}^{-2}
\label{eq:rho0}
\end{equation}
\citep[section 9 of][]{Gray1992, Suzuki2018} from the solar value, $\rho_{0,\odot}=2.5\times 10^{-7}$g cm$^{-3}$ adopted from the ATLAS model atmosphere \citep{Kuruz1979, Castelli2003}. 
As shown in Table \ref{tab:stprm}, $\rho_0$ of the RGB stars is lower than that of the Sun because the atmosphere is extended, causing the photosphere with the optical depth $\sim 1$ to form at a lower-density location. This explains the positive dependence 
of $\rho_0$ on $g$ in equation (\ref{eq:rho0}).
We note that the values of $\rho_0$ are consistent with, but slightly lower than, those derived from the ATLAS tabulation, $\rho_0\approx 1.3\times 10^{-8}$g cm$^{-3}$  for $\alpha$ Boo and $1.0\times 10^{-8}$ g cm$^{-3}$  for $\alpha$ Tau, respectively. The MARCS atmosphere model \citep{Gustafsson2008} also give similar results. 

We derive the fluctuation velocity, $\delta v_0$, at the photosphere from convective flux:
\begin{equation}
  \rho_0\delta v_0^3 \propto \rho_0\delta v_\mathrm{conv}^3= \frac{\alpha(\gamma -1)}{2\gamma}\sigma T_\mathrm{eff}^4 \propto T_\mathrm{eff}^4,
  \label{eq:dv0}
\end{equation}
where $v_\mathrm{conv}$ is convective velocity, $\alpha$ is the mixing length parameter and $\gamma$ is the ratio of the specific heats. 
To use this relation, we adopt the solar normalization, $\delta v_{0,\odot}=1.25$ km s$^{-1}$, which was used in previous simulations for solar winds \citep[][M24]{Suzuki2018}. This value is determined by observations of granular motion on the solar photosphere \citep[e.g.,][]{Matsumoto2010, Oba2020},  We note that $\delta v_0$ is smaller than $\delta v_\mathrm{conv}$ because the photosphere is generally located above the top of the convectively unstable region. The solar convective velocity can be estimated to be $\delta v_{\mathrm{conv},\odot}=4.2$ km s$^{-1}$ for $\alpha=1.5$ and $\gamma=5/3$ from equation (\ref{eq:dv0}). This shows $\rho_0\delta v_{0,\odot}^3\ll \rho_0 \delta_{\mathrm{conv},\odot}^3$, indicating that an only small fraction of the convective energy is utilized to generate MHD waves.  
    We determine $\delta v_0$ of the RGB stars in Table \ref{tab:stprm} by using a simple scaling $\delta v_0 \propto (T_\mathrm{eff}^4/\rho_0)^{1/3}$ with the solar normalization.

We assume the same amplitude for the longitudinal (normal to the surface) and two transverse (parallel with the surface) components of fluctuations in a frequency range between $\omega_\mathrm{min}$ and $\omega_\mathrm{max}=100\omega_\mathrm{min}$:
\begin{equation}
\langle \delta v_0^2\rangle = \int_{\omega_\mathrm{min}}^{\omega_\mathrm{max}}d\omega P(\omega), 
\end{equation}
where we adopt a power spectrum, $P(\omega)\propto \omega^{-1}$ \citep[][M24]{Suzuki2005}.
The typical period 
of the fluctuations can be scaled as follows \citep[][]{Suzuki2007,Airapetian2010,Suzuki2018}:
\begin{equation}
  \omega_\mathrm{min,max}^{-1} \propto \frac{H_0}{c_\mathrm{s,0}} = \frac{c_\mathrm{s,0}}{g}\propto \frac{c_\mathrm{s,0} R_{\star}^2}{M_{\star}},
  \label{eq:wvP}
\end{equation}
where $c_\mathrm{s,0}$ is the isothermal sound velocity and $H_0 = c_\mathrm{s,0}^2/g$ is the pressure scale height, both evaluated at the inner boundary in the photosphere.
The physical background of equation (\ref{eq:wvP}) is that $\omega^{-1}$ is proportional to the eddy turnover time, $\sim l/c_\mathrm{s,0}$, where the eddy size, $l$, is further scaled with $H_0$ \citep[][]{Stein2004,Suzuki2007}. We note that the scaling of equation (\ref{eq:wvP}) is the same as that of the acoustic cut-off frequency \citep{Samadi2007}.
As the normalization for the scaling, we adopt $\omega_\mathrm{max,\odot}^{-1}=0.3$ minutes and $\omega_\mathrm{min,\odot}^{-1}=30$ minutes used in our solar wind simulations \citep[][M24]{Suzuki2018}. The typical period of the RGB stars is much longer owing to the lower surface gravity, resulting in $\omega_\mathrm{max}^{-1}=150$ (340) minutes and $\omega_\mathrm{min}^{-1}=1.5\times 10^4$ ($3.4\times 10^4$) minutes for $\alpha$ Boo ($\alpha$ Tau) (Table \ref{tab:stprm}). 

\begin{table*}
  \tbl{Simulation models for $\alpha$ Boo.\footnotemark[$*$] }{
    \begin{tabular}{c|ccccccccc}
      \hline
      $f_0 B_{r,0}$ [G] & 0.65 & 0.65 & 0.37 & 1.31 & 0.65 & 0.65 & 0.65 & 0.65 & 0.65 \\
      Magnetic diffusion & On & Off & On & On & On & On & On & On & On \\      
      $Z (Z_{\odot})$ & 0.3$^{**}$ & 0.3$^{**}$ & 0.3$^{**}$ & 0.3$^{**}$ & 1 & $10^{-1}$ & $10^{-2}$ & $10^{-3}$ & 0 \\
      $N$ [number of cells] & 1800 & 3200 & 2500 & 1080 & 1800 & 1800 & 1800 & 1800 & 1800 \\
      \hline 
            $\langle{\dot{M}}\rangle$ [$M_{\odot}$yr$^{-1}$]$^{\dag}$ & $3.3(-11)$ & $5.1(-10)$ & $6.1(-12)$ & $4.0(-11)$ & $2.6(-11)$ & $1.9(-11)$ & $3.2(-12)$ & $3.5(-13)$ & $1.2(-14)$ \\
      $\dot{M}_\mathrm{min}$ [$M_{\odot}$yr$^{-1}$]$^{\dag \dag}$ & $1.0(-12)$ & $9.3(-12)$ & $2.8(-13)$ & $1.3(-12)$ & $3.2(-13)$ & $1.2(-12)$ & $3.0(-14)$ & $2.5(-14)$ & $7.1(-16)$ \\
      $\dot{M}_\mathrm{max}$ [$M_{\odot}$yr$^{-1}$]$^{\dag \dag}$ & $3.8(-10)$& $4.5(-9)$ & $6.2(-11)$ & $5.4(-10)$ & $3.7(-10)$ & $1.1(-10)$ & $2.2(-11)$ & $1.3(-12)$ & $2.0(-13)$ \\
      $\langle v_\mathrm{out}\rangle$ [km s$^{-1}$]$^{\dag}$ & 77 & 45 & 147 & 180 & 44 & 184 & 347 & 710 & 144 \\
      $\langle L_{\mathrm{0.5MK}-}\rangle$ [erg s$^{-1}$]$^\ddag$ & $7.9(26)$ &  $1.0(28)$ & $1.1(25)$ & $2.7(27)$ & $9.4(26)$ & $9.8(25)$ & $2.5(24)$ & $4.0(19)$ & $0$ \\
      $\langle L_{\mathrm{20 kK}-\mathrm{0.5MK}}\rangle$ [erg s$^{-1}$]$^\ddag$ & $4.0(29)$ & $1.3(30)$ & $9.3(28)$ & $6.1(29)$ & $5.1(29)$ & $2.8(29)$ & $6.1(28)$ & $6.8(27)$ & $8.9(26)$ \\
      \hline
    \end{tabular}}\label{tab:runsBoo}
  \begin{tabnote}
    \footnotemark[$*$] The upper half is input parameters and the lower half shows outputs of the simulations. ``$a(b)$'' in the outputs stands for $a\times 10^b$. \\
    \footnotemark[$**$] The heavy element abundances of the standard cases for $\alpha$ Boo are presented in Table \ref{tab:metal}.\\
     \footnotemark[$\dag$] $\langle \dot{M} \rangle$ and $\langle v_\mathrm{out}\rangle$ are respectively the mass-loss rate (equation \ref{eq:masslossrate}) and the terminal velocity averaged over time, $0.1t_\mathrm{sim}$ and $t_\mathrm{sim}$, (Section \ref{sec:numsetup}) and radial distance, $28R_{\star}$ and $31R_{\star}$ (Section \ref{sec:timevariation}).  \\
     \footnotemark[$\dag \dag$] $\dot{M}_\mathrm{max}$ and $\dot{M}_\mathrm{min}$ are the maximum and minimum mass-loss rates during this time duration. \\
     \footnotemark[$\ddag$]$\langle L_{T_1-T_2}\rangle$ is the time averaged radiation luminosity from the gas in $T_1<T<T_2$ (equation \ref{eq:Lrad}).     
  \end{tabnote}
\end{table*}

\begin{table}
  \tbl{Simulation models for $\alpha$ Tau.}{
    \begin{tabular}{c|ccc}
      \hline
      $f_0 B_{r,0}$ [G] & 0.25& 0.25 & 0.50 \\
      Magnetic diffusion & On & Off & On\\      
      $Z (Z_{\odot})$ & 0.74 & 0.74 & 0.74\\
      $N$ [number of cells] & 5300 & 10000 & 2800\\
      \hline 
            $\langle{\dot{M}}\rangle$ [$M_{\odot}$yr$^{-1}$] & $1.5(-12)$ & $7.0(-11)$ & $2.0(-11)$ \\
      $\dot{M}_\mathrm{min}$ [$M_{\odot}$yr$^{-1}$] & $9.9(-14)$ & $-1.0(-12)$ & $8.7(-13)$ \\
      $\dot{M}_\mathrm{max}$ [$M_{\odot}$yr$^{-1}$] & $1.4(-11)$ & $6.8(-10)$ & $3.1(-10)$\\
      $\langle v_\mathrm{out}\rangle$ [km s$^{-1}$] & 179 & 12 & 52 \\
      $\langle L_{\mathrm{0.5MK}-}\rangle$ [erg s$^{-1}$] & $0$ & $6.0(28)$ & $3.1(25)$\\
      $\langle L_{\mathrm{20 kK}-\mathrm{0.5MK}}\rangle$ [erg s$^{-1}$] & $7.2(27)$ & $2.7(30)$ & $1.9(29)$\\
      \hline
  \end{tabular}}
  \label{tab:runsTau}
\end{table}

\subsection{Magnetic Field}
\label{sec:Bfield}
The numerical simulations are performed in a super-radially open flux tube that is anchored at a small-scale patch with strong magnetic field, following our previous works for solar winds \citep[][M24]{Suzuki2005}. We assume the equipartition between the magnetic pressure, 
{$B_{r,0}^2/(8\pi)$}, and the gas pressure,
\begin{equation}
  p_0 = \frac{\rho_0}{\mu m_\mathrm{u}}k_\mathrm{B}T_\mathrm{eff} \equiv \rho_0 c_\mathrm{s,0}^2
  \label{eq:p0}
\end{equation}
at the photosphere: 
\begin{equation}
  \frac{8\pi p_0}{B_{r,0}^2} = 1,
  \label{eq:eqP}
\end{equation}
where $m_\mathrm{u}$ is the atomic mass unit and $k_\mathrm{B}$ is the Boltzmann constant; 
the mean molecular weight per particle, $\mu$, is determined by calculating ionization and recombination balance (equation \ref{eq:mu} in Section \ref{sec:ion}). 
The radial field strength, $B_{r,0}$, at the photosphere is derived from equation (\ref{eq:eqP}), which is 200-300 G for the RGB stars (Table \ref{tab:stprm}).
We note that these values are comparable to those employed in \citet{Suzuki2007}.

A sizable fraction of the surface is thought to be covered by magnetically closed loops. Then, the cross section, $A\propto r^2f$, of an open magnetic flux tube expands more rapidly than the simple radial expansion $\propto r^2$, where $r$ is the distance from the center of the star and $f$ is the filling factor, 
\begin{equation}
  f(r) = \frac{e^{\frac{r-R_{\star}-h}{\zeta}} + f_0 - (1-f_0)e^{-\frac{h}{\zeta}}}{e^{\frac{r-R_{\star}-h}{\zeta}} + 1}
  \label{eq:fillfac}
\end{equation}
\citep[][M24]{Suzuki2013}, where $h$ is a typical height of closed loops and $\zeta = h/2$. We assume that $h$ is proportional to the pressure scale height:
\begin{equation}
  \frac{h}{R_{\star}} \propto \frac{H}{R_{\star}}=\frac{c_\mathrm{s,0}^2}{R_{\star}g},
  \label{eq:h/R}
\end{equation}
where $h/R_\star$ for the simulated RGB stars in Table \ref{tab:stprm} are scaled from $h=0.042R_{\star}$ used in our solar wind simulations (M24).
The flux tube opens super-radially in $R_\star + h -\zeta \lesssim r \lesssim R_\star + h +\zeta$ and approaches to the radial expansion above $r\gtrsim R_\star + h + \zeta$.   
We note that the filling factor is proportional to the widely used super-radial expansion factor, $f_\mathrm{ex}$, \citep{Kopp1976,Suzuki2006ApJL} via $f(r) =f_0f_\mathrm{ex}(r)$.

The radial distribution of $B_r$ is determined by the conservation of magnetic flux, 
$\mbf{\nabla\cdot B} = 0$, as 
\begin{equation}
  B_r(r) = B_{r,0}\frac{R_{\star}^2 f_0}{r^2 f}.
  \label{eq:divB}
\end{equation}
\citet{Sennhauser2011} detected the longitudinal component of magnetic fields with $0.65\pm 0.26$ G and $0.43\pm 0.16$ G at different times in $\alpha$ Boo with their Zeeman component decomposition technique. \citet{Auriere2015} reported a slightly weaker field of $0.34\pm 0.11$ G, whereas the $S$-index based on chromospheric lines predicts 
a somewhat stronger field strength of $\sim 1$ G (Fig.10 of their paper; see Section \ref{sec:Tau_strong} for related discussion on $\alpha$ Tau).
The maximum and minimum values of these observational data with the errors are presented in $f_0B_{r,0}$ of Table \ref{tab:stprm}, which is the average field strength contributed from the open field area. 
This is because the observed longitudinal fields likely reflect the open field component rather than the closed structure in the disk center of a star, as closed loops are dominated by horizontal fields with respect to the line of sight from an observer.
Therefore, we can utilize these observational data to determine $f_0$.
We adopt $f_0B_{r,0}=0.65$ G, giving $f_0=1/403$, for our fiducial runs  of $\alpha$ Boo (Table \ref{tab:runsBoo}).  We also perform simulations with weaker and stronger fields to inspect the dependence on the magnetic field (Section \ref{sec:depB}).

Sub-G magnetic fields are also detected in $\alpha$ Tau \citep{Auriere2015}. Interestingly, the field polarity is reversed, depending on observational times: $0.22\pm 0.08$ G in March 2010, $-0.25\pm 0.13$ G in October 2010, and $0.22\pm 0.09$ G in January 2011.  The range of these strengths is shown in absolute values in Table~\ref{tab:stprm}. 
We take $f_0 B_{r,0}=0.25$ G, yielding $f_0=1/790$, in our simulations for $\alpha$ Tau.
Additionally, we also test a case with a field that is twice as strong
(Table \ref{tab:runsTau}), which will be discussed in Section~\ref{sec:Tau_strong}.

\subsection{MHD Equations}
\label{sec:MHDeqs}
We perform one-dimensional (1D) MHD simulations in spherical coordinates, explicitly taking into account magnetic diffusion, turbulent dissipation of Alfv\'en(ic) waves, radiative cooling, and thermal conduction.
We are solving the equations for mass continuity, 
\begin{equation}
  \label{eq:mass}
  \frac{d\rho}{dt} + \frac{\rho}{r^2 f}\frac{\partial}{\partial r}(r^2 f v_r ) = 0 , 
\end{equation}
radial momentum, 
\begin{align}
  \label{eq:mom}
  \rho \frac{d v_r}{dt} &= -\frac{\partial p}{\partial r}  
  - \frac{1}{8\pi r^2 f}\frac{\partial}{\partial r}  (r^2 f B_{\perp}^2) \nonumber \\
  &+ \frac{\rho v_{\perp}^2}{2r^2 f}\frac{\partial }{\partial r} (r^2 f)
  -\rho \frac{G M_{\star}}{r^2}  , 
\end{align}
transverse momentum, 
\begin{equation}
  \label{eq:moc1}
  \rho \frac{d}{dt}(r\sqrt{f} \mbf{v_{\perp}}) = \frac{B_r}{4 \pi} \frac{\partial} 
       {\partial r} (r \sqrt{f} \mbf{B_{\perp}}) + \rho \mbf{D}_{v_\perp}, 
\end{equation}
total energy, 
\begin{align}
  \label{eq:eng}
  \rho &\frac{d}{dt}\left(e + \frac{v^2}{2} + \frac{B^2}{8\pi\rho}
  - \frac{G M_{\star}}{r} \right) \nonumber \\
  &+ \frac{1}{r^2 f} \frac{\partial}{\partial r}\left[r^2 f \left\{ \left(p 
    + \frac{B^2}{8\pi}\right) v_r  
    - \frac{B_r}{4\pi} (\mbf{B \cdot v}) + F_\mathrm{c}\right\}\right]
   \nonumber \\
  &= \frac{1}{r^2f}\frac{\partial}{\partial r}\left[\frac{\eta_\mathrm{tot}}{4\pi}r\sqrt{f}\mbf{B}_{\perp} \frac{\partial}{\partial r}(r\sqrt{f}\mbf{B}_{\perp})\right] - Q_\mathrm{rad},
\end{align}
and transverse magnetic field, 
\begin{align}
  \label{eq:ct}
  \frac{\partial \mbf{B_{\perp}}}{\partial t} &= \frac{1}{r \sqrt{f}}
  \frac{\partial}{\partial r} [r \sqrt{f} (\mbf{v_{\perp}} B_r - v_r \mbf{B_{\perp}})] \nonumber \\
  &+ \sqrt{4\pi\rho}\mbf{D}_{B_\perp} + \frac{1}{r\sqrt{f}}\frac{\partial}{\partial r}
  \left[\eta_\mathrm{tot}\frac{\partial}{\partial r}(r\sqrt{f}\mbf{B_{\perp}})\right], 
\end{align}
where 
{$d/dt$ and $\partial/\partial t$} stand for the Lagrangian and Eulerian time derivatives, respectively; 
$\rho$, $\mbf{v}$, $\mbf{B}$, $p$, and  $e$ are density, velocity, magnetic field, gas pressure, and specific energy, respectively, where 
\begin{equation}
  e=\frac{p}{\rho (\gamma -1)} = \frac{k_\mathrm{B}T}{\mu m_\mathrm{u}(\gamma -1)} .
  \label{eq:e2T}
\end{equation}
For $\mbf{v}$ and $\mbf{B}$, in addition to the radial component, we consider the two transverse components, e.g., $\mbf{v}_{\perp}= (v_{\perp 1}, v_{\perp 2})$ and $v_{\perp}^2\equiv v_{\perp 1}^2+v_{\perp 2}^2$. 
{$F_\mathrm{c} = -\kappa_0 T^{\frac{5}{2}}\partial T/ \partial r$} is thermal conduction flux through Coulomb collisions by electrons, where $\kappa_0=10^{-6}$erg cm$^{-1}$s$^{-1}$K$^{-7/2}$ \citep{Braginskii1965}. We note that the conduction is important in the fully ionized high-temperature gas with $T > 10^5$K but is negligible in the weakly ionized gas with $T\lesssim 10^4$K.
$\mbf{D}_{v_\perp}$ and $\mbf{D}_{B_\perp}$ respectively denote turbulent dissipation of velocity and magnetic field fluctuations of Alfv\'enic waves (Section \ref{sec:turb}). $\eta_\mathrm{tot}=\eta_\mathrm{O} + \eta_\mathrm{AD}$ is the sum of Ohmic resistivity, $\eta_\mathrm{O}$, and ambipolar diffusivity, $\eta_\mathrm{AD}$, which will be described in Section \ref{sec:Bdif}. 

$Q_\mathrm{rad}$[erg cm$^{-3} {\rm s}^{-1}$] represents radiative cooling.
In the high-temperature gas with $T>1.2\times 10^4$K, we employ an optically thin cooling for ionized gas:
\begin{equation}
  Q_\mathrm{rad} = n_i n_e \Gamma,
  \label{eq:Qrad_hT}
\end{equation}
where $n_i$ and $n_e$ are the ion and electron number densities, respectively, and $\Gamma$ is the radiative loss function taken from \citet{Sutherland1993}.
In the low-temperature gas with $T\le 1.2\times 10^4$K, we use an empirical cooling based on observations of the solar chromosphere \citep{Anderson1989} with modification to consider the dependence on metallicity \citep{Suzuki2018} and the effect of the optically thin and thick transition (M24):
\begin{equation}
  Q_\mathrm{rad} = 4.5\times 10^9\left(0.2+0.8\frac{Z}{Z_\odot}\right)\min\left(1,\frac{\rho}{\rho_\mathrm{cr}}\right)
  {\mathrm{[erg\;cm^{-3} s^{-1}]}},\label{eq:Qrad_lT}
\end{equation}
where $\rho_\mathrm{cr}=10^{-16}$g cm$^{-3}$.
We switch off the radiative cooling below $T=T_\mathrm{cut} =0.7T_\mathrm{eff}$ ($3010$ K for $\alpha$ Boo and $2730$ K for $\alpha$ Tau) for numerical stability. 

\subsection{Turbulent Dissipation}
\label{sec:turb}
We phenomenologically take into account the dissipation of Alfv\'enic waves by turbulent cascade \citep[e.g.,][]{Hossain1995,Matthaeus1999,Cranmer2005}. We follow the formalism introduced by \citet{Shoda2018a}, which gives
\begin{equation}
  D_{v_{\perp i}} = -\frac{c_\mathrm{d}}{4\lambda_{\perp}}(|z_{\perp i}^{+}|z_{\perp i}^{-} + |z_{\perp i}^{-}|z_{\perp i}^{+} )
\end{equation}
and
\begin{equation}
  D_{B_{\perp i}} = -\frac{c_\mathrm{d}}{4\lambda_{\perp}}(|z_{\perp i}^{+}|z_{\perp i}^{-} - |z_{\perp i}^{-}|z_{\perp i}^{+} ), 
\end{equation}
where 
\begin{equation}
   \mbf{z}_\perp^{\pm} = \mbf{v} \mp \frac{\mbf{B}_\perp}{\sqrt{4\pi\rho}}
  \label{eq:zpp}
\end{equation}
is the Els\"asser variables \citep{Elsasser1950}, and $i=1,2$. We set the dimensionless constant, $c_\mathrm{d}=0.1$ \citep{vanBallegooijen2017} and $\lambda_{\perp}$ is the correlation length that is assumed to be proportional to the horizontal scale of the flux tube,
\begin{equation}
  \lambda_{\perp}(r) = \lambda_{\perp 0}\sqrt{\frac{r^2f(r)}{R_{\star}^2 f_0}},   
\end{equation}
where we estimate $\lambda_{\perp 0}$ at the photosphere by an extension from the solar condition. 
There is still large uncertainty in observationally inferred solar values from a sub-granular scale, $\lambda_{\perp 0 \odot}\approx 300$ km \citep{Cranmer2005,Cranmer2011}, to a supergranular scale, $\lambda_{\perp 0 \odot}\approx (7.6-9.3)\times 10^3$ km \citep{Sharma2023}. We employ an intermediate value, $\lambda_{\perp 0 \odot}=3\times 10^3$ km, which lies between the sizes of granular and supergranular cells, for the solar normalization.
The typical scale of granular/supergranular cells is regarded to be proportional to the pressure scale height \citep{Freytag2002, Magic2013}.
Equation (\ref{eq:h/R}) gives the scale height of $\alpha$ Boo as 330 times the solar scale height. Using this value, we adopt $\lambda_{\perp 0}=1.0 \times 10^6$ km for $\alpha$ Boo. The same procedure gives $\lambda_{\perp 0}=2.2 \times 10^6$ km for $\alpha$ Tau.

\subsection{Ionization}
\label{sec:ion}
It is necessary to determine the ion and neutral populations in order to calculate magnetic diffusion (see Section \ref{sec:Bdif}).
We define ``the ionization degree'':
\begin{equation}
  x_e = \frac{n_e}{n_n+ n_i}, 
\end{equation}
where 
$n_n$ is the number density of neutral particles\footnote{The ionizatioin degree is usually defined as 
{$n_i/(n_n +n_i)$}. If the electrons are supplied from singly charged ions, $x_e$ is the same as the this definition, but if part of the electrons come from ions that are more than singly ionized, $x_e$ is larger than the original definition. }.
Although in M24 $x_e$ was defined against the number density of hydrogen, we modified the normalization to all the elements because the improved definition is more directly connected to the diffusion coefficients as described later in Section \ref{sec:Bdif}. From the charge neutrality, we obtain 
\begin{equation}
  n_e = n_{\mathrm{H}^+} + n_{\mathrm{He}^{+}} + 2n_{\mathrm{He}^{++}} + \sum_j n_{j^+}, 
\end{equation}
where the right-hand side indicates the sum of the number densities of respective ions. The last term denotes the contribution from the heavy elements with low FIP (Section \ref{sec:abnd} and Table \ref{tab:metal}); we consider only singly charged ions of these elements because, when they are doubly ionized, the dominant portion of the electrons are already supplied from the ionization of hydrogen. 

We adopt the same procedure for the ionization of H and He as in M24; we adopt an analytic model for the excitation and ionization of H introduced by \citet[][see also \citet{Harper2001}]{Hartmann1984} and the fractions of He, He$^+$, and He$^{2+}$ are determined under the local thermodynamic equilibrium (LTE).
We calculate the population ratio of the ions and neutrals in the $j$-th heavy element as
\begin{align}
  \frac{n_{j^+}}{n_{j^0}} &= \frac{1}{n_e}\frac{2 g^+}{g^0}\left(\frac{2\pi m_e k_\mathrm{B}}{h^2}\right)^{\frac{3}{2}}\sqrt{T} \nonumber \\
  &\times \max\left[W T_\mathrm{eff} e^{-\frac{E_\mathrm{i}}{k_\mathrm{B}T_\mathrm{eff}}}  \right. \nonumber \\
    &\left. + W_\mathrm{gal} T_\mathrm{gal} e^{-\frac{E_\mathrm{i}}{k_\mathrm{B}T_\mathrm{gal}}}, Te^{-\frac{E_\mathrm{i}}{k_\mathrm{B}T}} \right],
  \label{eq:Saha}
\end{align}
where $g^+$ and $g^0$ are the statistical weights of the ion and the neutral, respectively, $E_\mathrm{i}$ is the ionization energy, and the elemental abundance is normalized as $n_{j^0} + n_{j^+} = A_j(n_{\mathrm{H}^0} + n_{\mathrm{H}^{+}} )$.
$W = 1/2 [ 1-\sqrt{1-\left(R_\star/r\right)^2}]$ is the geometric dilution factor and the term with $W_\mathrm{gal}=10^{-14}$ and $T_\mathrm{gal}=7500$ K represents the contribution from the interstellar radiation field \citep{Mathis1983}. Although M24 considered the only radiative ionization by the central star (term with $T_\mathrm{eff}$) and the interstellar radiation field (term with $T_\mathrm{gal}$), in the current work we additionally take into account thermal ionization, which possibly dominates in transiently formed high-temperature regions (Section \ref{sec:timevariation});  we include this process in a simplified manner, instead of simultaneously solving both radiative and thermal ionization/recombination balance, as the second component of the $\max$ function, which corresponds to the solution of the Saha's equation under the LTE. 

Once $x_e$ is determined, we can calculate the mean molecular weight per particle,
\begin{equation}
  \mu = \frac{1+4A(\mathrm{He}) + \sum_jN_jA_j}{(1+A(\mathrm{He}) + \sum_jA_j)(1+x_e)}.
  \label{eq:mu}
\end{equation}
which 
regulates the relation between specific energy and temperature and determines gas pressure from density and temperature (equation \ref{eq:e2T}). 
We should note that the current treatment for the equation of state and $\mu$ is still incomplete and needs further elaboration. First, molecules have to be considered in low-temperature regions, which will give larger $\mu$. 
Second, the latent heats by ionization/recombination and formation/dissociation of molecules need to be included, in addition to the thermal specific energy, in equation (\ref{eq:e2T}) \citep[e.g.,][]{vardya1965,Iijima2016}. The ignorance of the latent heat could overestimate temperature change; in particular temperature decrease 
by the 
radiative cooling, 
mainly because the recombination is overestimated with respect to the ionization (equations \ref{eq:Qrad_hT} and \ref{eq:Qrad_lT}).
On the other hand, the artificial cut-off of the radiation cooling for $T<T_\mathrm{cut}$ (Section \ref{sec:MHDeqs}) offsets this overcooling in the low-temperature range. The effect of molecules and latent heats should be included in our more elaborated future studies. 

\subsection{Magnetic Diffusion}
\label{sec:Bdif}
The diffusivity by Ohmic resistivity is calculated from the collision between electrons and neutrals \citep{Blaes1994},
\begin{align}
  \eta_\mathrm{O} &= \frac{c^2m_e \nu_{en}}{4\pi e_c^2n_e} \nonumber \\
  &\approx 2.3\times 10^{2}\frac{\max(1-x_e,0)}{x_e}\sqrt{T\;\mathrm{[K]}}\; \mathrm{[cm^2 s^{-1}]},
  \label{eq:etaO}
\end{align}
where $c$ is the speed of light, $e_c$ is the elementary charge, and $m_e$ is the mass of an electron. The numerical value of the electron-neutral collision frequency, $\nu_{en}$, is adopted from \citet{Draine1983}. The ambipolar diffusivity for the collision between ions and neutrals is approximately given by\footnote{In M24, a square is applied to the $\max(1-x_e,0)$ term, but the presented expression is correct. Strictly speaking, $x_i\equiv n_i/(n_i+n_n)$ should be used instead of $x_e$ as $\rho_n/\rho_i = (1-x_i)/x_i$. $x_e$ is slightly larger than 1 in the fully ionized condition, and hence, the cap by the $\max$ function is applied. We note, however, that the difference between $x_e$ and $x_i$ is tiny because it is only due to the contribution from doubly ionized He in our setup.} 
\begin{align}
  \eta_\mathrm{AD} &= \frac{B^2(\rho_n / \rho)^2}{4\pi \chi \rho_i\rho_n}\nonumber \\
  &\approx 2.1\times 10^{-16}\frac{\max(1-x_e,0)}{x_e }\left(\frac{B\;{\mathrm{[G]}}}{\rho\mathrm{[g\; cm^{-3}]}}\right)^2\; \mathrm{[cm^2 s^{-1}]}
  \label{eq:etaAD}
\end{align}
\citep[Chapter 27 of][]{Shu1992,Khomenko2012}, where $\rho_i$ and $\rho_n$ are the mass densities of ions and neutrals, respectively.
$\chi$ is a term arising from ion-neutral collisions (M24), with its numerical value 
adopted from \citet{Draine1983}. 

For the analyses of magnetic diffusion processes, we introduce the Magnetic Reynolds number,
\begin{equation}
  \mathrm{Rm_{tot}} = \frac{\Lambda V}{\eta_\mathrm{tot}} = \frac{\Lambda V}{\eta_\mathrm{O}+ \eta_\mathrm{AD}},
  \label{eq:Rm}
\end{equation}
where $\Lambda$ and $V$ are typical scale and velocity that estimate the order of the inertial term; we adopt the scale height, $\Lambda=H_0$ (equation \ref{eq:h/R}), and the sound velocity, $V=c_\mathrm{s,0}$ (equation \ref{eq:p0}), at the photosphere.

\subsection{Numerical Setup}
\label{sec:numsetup}
The simulations are carried out in a single magnetic flux tube that is super-radially open according to equation (\ref{eq:fillfac}). In reality, the stellar surface is covered by numerous flux tubes. For later discussions, we estimate the number, $\cal{N}$, of open flux tubes that covers the total surface multiplied by the filling factor, $4\pi R_{\star}^2 f_0$. If we define $d_0$ as the diameter of a tube at the photosphere, we obtain ${\cal N}=4\pi R_{\star}^2 f_0/\pi (d_0/2)^2 = 16 \pi R_{\star}^2 f_0/d_0^2$. On the solar surface, magnetic flux tubes are anchored in intergranular lanes with typical widths of 100 -- 300 km \citep[e.g.,][]{Nordlund2009,Oba2017}, and we take $d_{0,\odot}=200$ km as a representative value, giving ${\cal N}_{\odot}\approx 1.5\times 10^5$ for the solar value, where $f_{0, \odot}=1/1265$ (M24). This value is consistent with the number of flux tubes measured at 1au: $4\pi r_\mathrm{1au}^2/\pi(d_\mathrm{1au}/2)^2 = 1.8\times 10^5$, where we adopt $d_\mathrm{1au}=1.4\times 10^6$ km for the mean diameter size of the flux tubes \citep{Borovsky2008}. Using the same scaling as the scale height, $d_0\propto h \propto c_{\mathrm{s},0}^2/g$ (equation \ref{eq:h/R}), 
for $d_0$, we have ${\cal N} \approx 2800$ and 950 for $\alpha$ Boo and $\alpha$ Tau, respectively, with the standard field strength presented in Tables \ref{tab:runsBoo} and \ref{tab:runsTau}. 


The simulated cases for $\alpha$ Boo and $\alpha$ Tau are listed in Tables \ref{tab:runsBoo} and \ref{tab:runsTau}. The leftmost case for each object is the standard nonideal MHD runs with the relevant magnetic field and metallicity. In order to 
{examine the roles of the} nonideal MHD effects, we also carry out ideal MHD simulations for comparison, indicated with ``B diff.: Off''. 
In addition, 
we further perform nonideal MHD simulations with different magnetic field strengths 
{($\alpha$ Boo and $\alpha$ Tau)} and metallicities ($\alpha$ Boo).

We dynamically solve the MHD equations in Section \ref{sec:MHDeqs} with the second order MHD Godunov-MoC method \citep{Sano1999}. At each time step, the Ohmic resistivity and ambipolar diffusivity (Section \ref{sec:Bdif}) are calculated with the ionization and recombination balance (Section \ref{sec:ion}). 
The simulation domain covers from the photosphere at $r=R_{\star}$ (Section \ref{sec:inbd}) to the outer boundary at $r=31R_{\star}$.
The size, $\Delta r$, of the numerical cells is arranged in order to resolve the Alfv\'en waves with 
{the shortest wavelength $v_\mathrm{A}/\omega_\mathrm{max}$ with at least four grid points at any $r$, where $v_\mathrm{A}=B_r/\sqrt{4\pi\rho}$ is the Alfv\'en velocity.}

In the fiducial case for $\alpha$ Boo, $\Delta r=5\times 10^{-4}R_\star (=8.8\times 10^3\;\mathrm{km})$ in the region close to the surface, and $\Delta r = 2\times 10^{-2}R_\star (=3.5\times 10^5\;\mathrm{km})$ in the outer region. We set the number of the cells, $N=1800$, for this case; $N$ of different runs is tabulated in Tables \ref{tab:runsBoo} and \ref{tab:runsTau}.
Since the wavelength is proportional to $v_\mathrm{A}$, cases with weaker magnetic field strength, $f_0 B_{r,0}$, and/or higher density, and accordingly larger mass-loss rate,
\begin{equation}
  \dot{M}=4\pi\rho v_r r^2 f,
  \label{eq:masslossrate}
\end{equation}
require smaller $\Delta r$ and a larger $N$. The ideal MHD run needs a larger $N$ than the corresponding nonideal MHD run because it yields  larger $\dot{M}$. 
As $\dot{M}$ is an output of the simulations, $N$ has to be determined through trial-and-error to meet the required accuracy mentioned above. 

We run the simulations until $t_\mathrm{sim} = 30R_{\star}/c_\mathrm{s,0}=1135$ [days] ($\approx 3.1$ [yrs]) and $=2144$ [days] ($\approx 5.9$ [yrs]) for $\alpha$ Boo and $\alpha$ Tau, respectively. We compare time-averaged profiles of various physical quantities of different runs later in Sections \ref{sec:massloss}, \ref{sec:Boo}, and \ref{sec:Tau}. The time average is taken between $0.1t_\mathrm{sim}$ and $t_\mathrm{sim}$ after outflowing structures are realized in all runs, whereas large time-variability is seen in most cases throughout the simulation time (Section \ref{sec:timevariation}).
We represent $\langle A\rangle$ as the time average of the variable $A$ over this duration in the following sections. 


\section{Mass-Loss Rates}
\label{sec:massloss}
\begin{figure}
  \begin{center}
    \includegraphics[height=6.3cm]{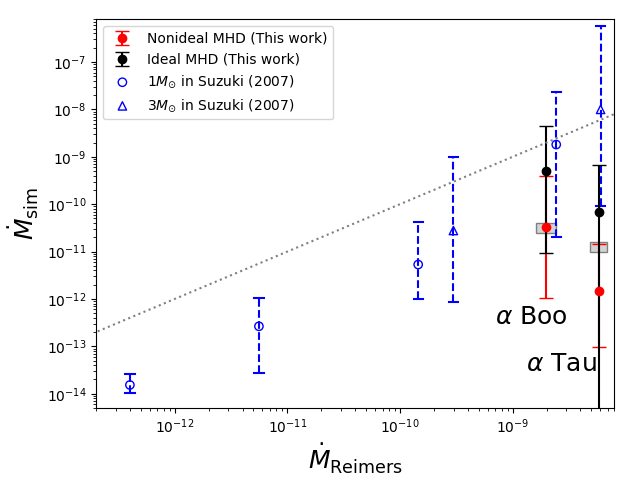}
  \end{center}
  \caption{Comparison of the mass-loss rates, $\dot{M}_\mathrm{sim}$, derived from numerical simulations with $\dot{M}_\mathrm{Reimers}$ empirically determined by the Reimers' formula, equation (\ref{eq:Reimers}). The symbols represent the time-averaged values and the 
    {vertical} bars indicate the range from the minimum to the maximum values of $\dot{M}$, which vary over time. The red and black filled circles denote the results of the nonideal and ideal MHD simulations, respectively. The open blue circles and triangles are taken from the ideal MHD simulations by \citet{Suzuki2007} for $M_\star=1M_\odot$ and $3M_\odot$, respectively.
  The shaded boxes are the observationally estimated values for $\alpha$ Boo \citep{Harper2022} and $\alpha$ Tau \citep{Robinson1998,Wood2007}. The dotted line shows $\dot{M}_\mathrm{sim} = \dot{M}_\mathrm{Reimers}$. 
  {Alt text: The horizontal range for $\dot{M}_\mathrm{Reimers}$ covers from $2\times 10^{-13}M_\odot$yr$^{-1}$ to $7\times 10^{-9}M_{\odot}$yr$^{-1}$ and the vertical range for $\dot{M}_\mathrm{sim}$ covers from $5 \times 10^{-15}M_\odot$yr$^{-1}$ to $8\times 10^{-7}M_{\odot}$yr$^{-1}$.}
  \label{fig:Mdotsum}}
\end{figure}

Figure \ref{fig:Mdotsum} compares the simulated mass-loss rates (vertical axis) with those empirically derived from the Reimers' relation (horizontal axis) \citep{Reimers1975},
\begin{equation}
  \dot{M}_\mathrm{Reimers} = 4\times 10^{-13} \left(\frac{R_\star}{R_\odot}\right)\left(\frac{L_\star}{L_\odot}\right)\left(\frac{M_\star}{M_\odot}\right)^{-1} M_\odot\;\mathrm{yr}^{-1}, 
  \label{eq:Reimers}
\end{equation}
where the simulated $\dot{M}_\mathrm{sim}$ in this study is evaluated by the spatial average of equation (\ref{eq:masslossrate}) between $r=28R_\star$ and $31R_\star$ just inside the outer boundary.
In addition to the results of this study for $\alpha$ Boo and $\alpha$ Tau (filled symbols),  the previous results from the MS to the RGB with ideal MHD simulations in \citet{Suzuki2007} (blue open symbols) are also presented.
The comparison between the horizontal and vertical axes shows that $\dot{M}_\mathrm{Reimers}$ 
 exceeds the observed mass-loss rates (shaded boxes) and most of the time-averaged $\dot{M}_\mathrm{sim}$ (symbols)\footnote{Later works \citep[e.g.][]{Michalitsianos1978,Migio2012}, including \citet{Reimers1977} himself, argue that a smaller value $\sim 10^{-13} M_\odot\;\mathrm{yr}^{-1}$ for the normalization constant is favored, which slightly reduces $\dot{M}_\mathrm{Reimers}$ \citep[see also][for modified mass-loss formulae]{Schroeder2005,Cranmer2011}.}. 

One can find that the result of the present study with the ideal MHD simulations (black) for $\alpha$ Boo reasonably follow the trend of \citet{Suzuki2007} although the detailed setting on the magnetic flux tube and surface fluctuations is slightly different.  
On the other hand,  $\dot{M}_\mathrm{sim}$ of the ideal MHD case for $\alpha$ Tau is smaller than expected from the trend because the weaker magnetic field strength is adopted (Table \ref{tab:runsTau}).
The nonideal MHD simulations give considerably smaller $\dot{M}_\mathrm{sim}$; the time-averaged $\langle \dot{M}_\mathrm{sim}\rangle \approx 3.3\times 10^{-11}M_{\odot}$yr$^{-1}$ for $\alpha$ Boo, which is about 1/15 of the time-averaged $\langle\dot{M}_\mathrm{sim}\rangle\approx 5.1\times 10^{-10}M_{\odot}$yr$^{-1}$ obtained in the ideal MHD counterpart, nicely explains the observed $\dot{M}=(2.5-4.0)\times 10^{-11}M_{\odot}$yr$^{-1}$ \citep[][shaded box in Figure \ref{fig:Mdotsum}]{Harper2022}.
$\dot{M}_\mathrm{sim}$ for $\alpha$ Tau is further reduced by 
{1.7 dex} from the ideal MHD setting to the nonideal setting. 
This is due to the lower ionization degree in the atmosphere, which enhances the ambipolar diffusion of Alfv\'enic waves, as will be discussed in the following Sections.   

Large time variability ranging about 2--3 orders of magnitude is seen in the nonideal MHD simulations as well as the ideal MHD simulations. This is due to transiently formed magnetized hot bubbles in the atmosphere, as will be discussed in Section \ref{sec:rprof}.
The multi-temperature wind structures may also affect the comparison of the mass-loss rates discussed above because the observational $\dot{M}$ is determined from the wind material with temperature, $T=(1-2)\times 10^4$K \citep{Harper2022}, which will also be discussed in Section \ref{sec:obsimp}.

Our simulations are conducted in a single magnetic flux tube. In reality, the total mass-loss rate is derived from the integration over the mass fluxes in numerous individual flux tubes that cover the stellar surface. Therefore, the time-variability in total integrated $\dot{M}$ is expected to be reduced by averaging the contributions from different flux tubes. As described in Section \ref{sec:method}, ${\cal N} \approx 2800$ flux tubes are distributed on the surface of $\alpha$ Boo. In the nonideal MHD case, the mass-loss rate in each flux tube varies in a range of +1.0-1.5 dex around the time-averaged value of $\log \langle\dot{M}[M_\odot\;\mathrm{yr}^{-1}]\rangle = -10.5$ (Table \ref{tab:runsBoo}), 
whereas the standard deviation can be also derived as $\approx 0.6$ dex by fitting the time sequence of snapshot $\dot{M}$ to a log-normal distribution (see Section \ref{sec:timevariation} for the time evolution of $\dot{M}$). Assuming that the flux tubes on the half hemisphere can be observed from us for simplicity, we obtain the standard deviation $=0.6 / \sqrt{{\cal N}/2}\approx 0.016$ dex of the $\dot{M}$ integrated over these randomly behaving flux tubes; namely  the typical time variability of the $\dot{M}$ integrated over the half surface is within $\pm 4$\%. Applying the same argument to $\alpha$ Tau, we can also derive the time variability of the integrated $\dot{M}$ is $\lesssim \pm 10$\%. 



\section{Arcturus}
\label{sec:Boo}
\subsection{Effects of Magnetic Diffusion}

\begin{figure}
  \begin{center}
    \hspace{-1cm}\includegraphics[height=12.3cm]{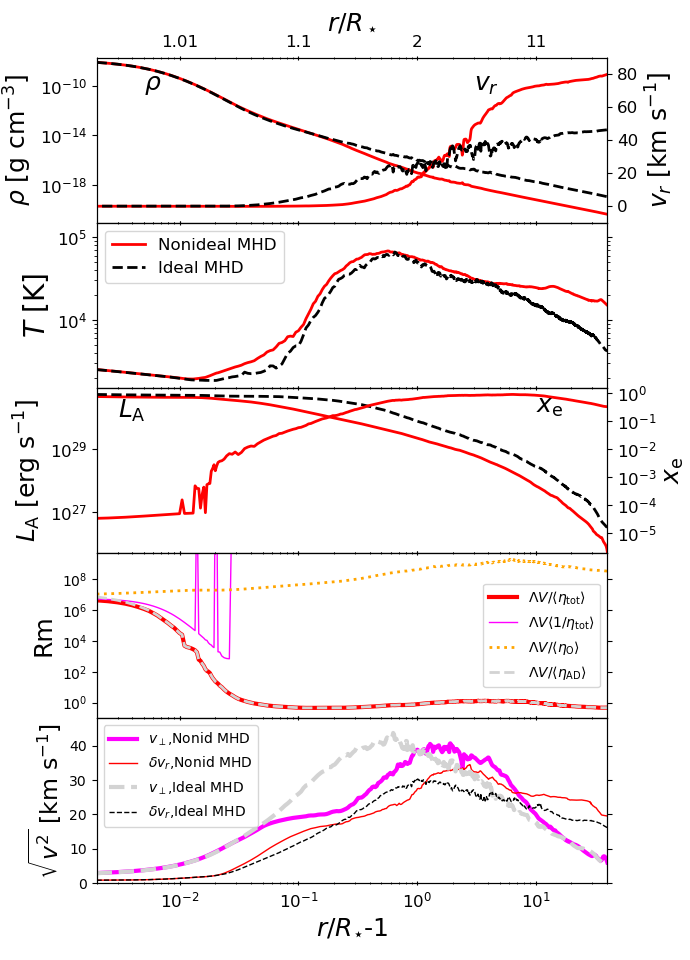}
  \end{center}
  \caption{Comparison of the 
    time-averaged radial profiles of various physical quantities between the nonideal MHD simulation (red solid lines in the top three panels) and the ideal MHD simulation (black dashed lines in the top three panels) for $\alpha$ Boo. The horizontal axis is the distance from the photosphere, $r-R_{\star}$, normalized by the stellar radius, $R_\star$ (bottom axis). For reference, the tick labels measured in $r/R_{\star}$ are also shown on the top axis. The top panel shows the densities (left axis) and radial velocities (right axis). The second panel compares the temperatures. The third panel presents the Alfv\'enic Poynting luminosities of both cases (left axis) and the ionization degree of the nonideal MHD case (right axis). The 
        fourth panel compares magnetic Reynolds numbers obtained from the nonideal MHD case. The thick red solid, gray dashed and orange dotted lines indicate those evaluated with the arithmetic time-averaged $\eta_\mathrm{tot}$, $\eta_\mathrm{AD}$ and $\eta_\mathrm{O}$, respectively. The thin pink solid line denotes $\langle \mathrm{Rm_{tot}}\rangle_\mathrm{hm}$ calculated with the harmonic time-average of the total magnetic diffusivity, $\Lambda V \langle 1/\eta_\mathrm{tot}\rangle$. The bottom panel compares rms transverse velocities (thick) and longitudinal velocity fluctuations (thin) of the nonideal (pink/red) and ideal (gray/black) cases. 
    {Alt text: Five panels from top to bottom respectively present four line, two line, three line, four line and four line graphs.}
  \label{fig:Non-ideal}}
\end{figure}

Figure \ref{fig:Non-ideal} compares the radial distributions of time-averaged various physical quantities 
of the nonideal (red solid) and ideal (black dashed) MHD simulations for $\alpha$ Boo as described in the caption. The top panel indicates that the density is lower in the nonideal MHD case in $r \gtrsim 1.2R_{\star}$, leading to the lower mass-loss rate, as stated above. Accordingly, the wind velocity is moderately larger as the lower-density gas can be accelerated more efficiently.  The final wind velocity of the nonideal case is 70-80 km s$^{-1}$ (Table \ref{tab:runsBoo}), which is slightly larger than the 
semi-empirical estimate $\sim 50$ km s$^{-1}$ by \citet{Harper2022} and 40-50 km s$^{-1}$ by \citet{Ayres1982} and \citet{Wood2016}. 

The second panel shows that both cases exhibit warm atmospheres with the peak temperature of several $10^4$ K at $r\approx 2R_{\star}$, which is moderately higher than the temperature of 15000 K estimated observationally by \citet{Harper2022}. However, we should cautiously note that these are merely the time-averaged profiles; the atmospheric temperature in both cases goes drastically up and down from $\lesssim 10^3$~K to $\gtrsim 10^6$~K as shown in Section \ref{sec:timevariation}. 

The ionization degree in the low atmosphere is low $\approx 10^{-5}-10^{-4}$  (third panel) because H and He are not ionized owing to the too low temperature and the electrons are supplied only from elements with low FIP. $\langle x_e\rangle$ gradually increases with height so that the gas is almost fully ionized, $\langle x_e\rangle\approx 1$, above $r\gtrsim 2 R_{\star}$.
The ideal MHD case also gives a similar $\langle x_e\rangle$ profile although it is not presented as the ionization degree does not substantially affect the ideal MHD simulation.
As a result of the weak ionization, the magnetic Reynolds number for the total diffusivity (thick red solid line in the fourth panel) is kept low, 
{$\langle\mathrm{Rm_{tot}}\rangle=\Lambda V/\langle\eta_\mathrm{tot}\rangle\sim 1$} 
in $r\gtrsim 1.05R_{\star}$. The comparison between the magnetic Reynolds numbers for Ohmic resistivity (orange dotted) and ambipolar diffusion (gray dashed) illustrates that the latter dominates in the entire region. 
These results indicate that the propagation of Alfv\'enic waves is greatly affected by ambipolar diffusion.  Indeed, the outgoing Alfv\'enic Poynting luminosity \citep{Suzuki2013,Shimizu2022},
\begin{equation}
  L_\mathrm{A} = \left[-B_r\frac{B_\perp v_\perp}{4\pi} + v_r\left(\rho\frac{v_\perp^2}{2} + \frac{B_\perp^2}{4\pi}\right)\right]4\pi \rho r^2 f,
  \label{eq:LA}
\end{equation}
in the nonideal MHD case is significantly reduced, compared with in the ideal MHD case,  before reaching the wind onset region. 
Consequently, the $\langle \dot{M}\rangle$ of the nonideal MHD case is greatly suppressed as explained in Figure \ref{fig:Mdotsum}.  

However, there is an important caveat in the discussion regarding the time-averaged magnetic Reynolds numbers. The magnetic Reynolds number, 
{$\langle \mathrm{Rm_{tot}} \rangle_\mathrm{hm}=\Lambda V\langle 1/\eta_\mathrm{tot}\rangle$}, calculated with the harmonic time-average of 
$\eta_\mathrm{tot}$ (thin pink solid line in the 
fourth panel of Figure \ref{fig:Non-ideal}) is much larger than that adopting the arithmetic average of 
$\eta_\mathrm{tot}$ (thick red solid line) except in the low-altitude region of $r<1.01R_{\star}$. In particular, $\langle \mathrm{Rm_{tot}} \rangle_\mathrm{hm}\rightarrow \infty$ above $r\gtrsim 1.05R_{\star}$, which is in clear contrast to 
{$\Lambda V/\langle\eta_\mathrm{tot} \rangle\sim 1$} in the same region. Since the harmonic mean is more sensitive to smaller values of $\eta_\mathrm{tot}$, the huge deviation between these two different time averages implies that $\eta_\mathrm{tot}$ temporally becomes $\Rightarrow 0$, which is discussed below in Section \ref{sec:timevariation}. 
 
The bottom panel of Figure \ref{fig:Non-ideal} compares the transverse and longitudinal components of root-mean-squared (rms) velocity fluctuations of the two cases. Both components are derived by density-weighted time integration:
\begin{equation}
\left\langle \sqrt{v_\perp^2}\right\rangle\equiv \sqrt{\frac{\int dt \rho (v_{\perp,1}^2+v_{\perp,2}^2)}{\int dt \rho}}
\end{equation}
and
\begin{equation}
\left\langle \sqrt{\delta v_r^2}\right\rangle\equiv \sqrt{\frac{\int dt \rho (v_r-\langle v_r\rangle)^2}{\int dt \rho}},
\end{equation}
where $\langle v_r\rangle=\int dt \rho v_r / \int dt \rho$ is also the density-weighted time average. Comparison of the transverse velocities (thick lines) of the two cases illustrates that $\langle v_\perp^2\rangle$ of the nonideal MHD case (pink thick solid line) is relatively suppressed in $r\gtrsim 1.05 R_\star$ because the Alfv\'enic waves are damped via ambipolar diffusion.  The longitudinal velocity fluctuations, which are mainly excited through parametric decay instability  from the transverse component\citep{Suzuki2006JGRA, Shoda2018b}, of the nonideal MHD case (thin red solid line) also follows a similar trend at a slightly higher location.  


\subsection{Time Variability}
\label{sec:timevariation}

\subsubsection{Wind and Radiation Properties}
\begin{figure}
  \begin{center}
    \includegraphics[height=11cm]{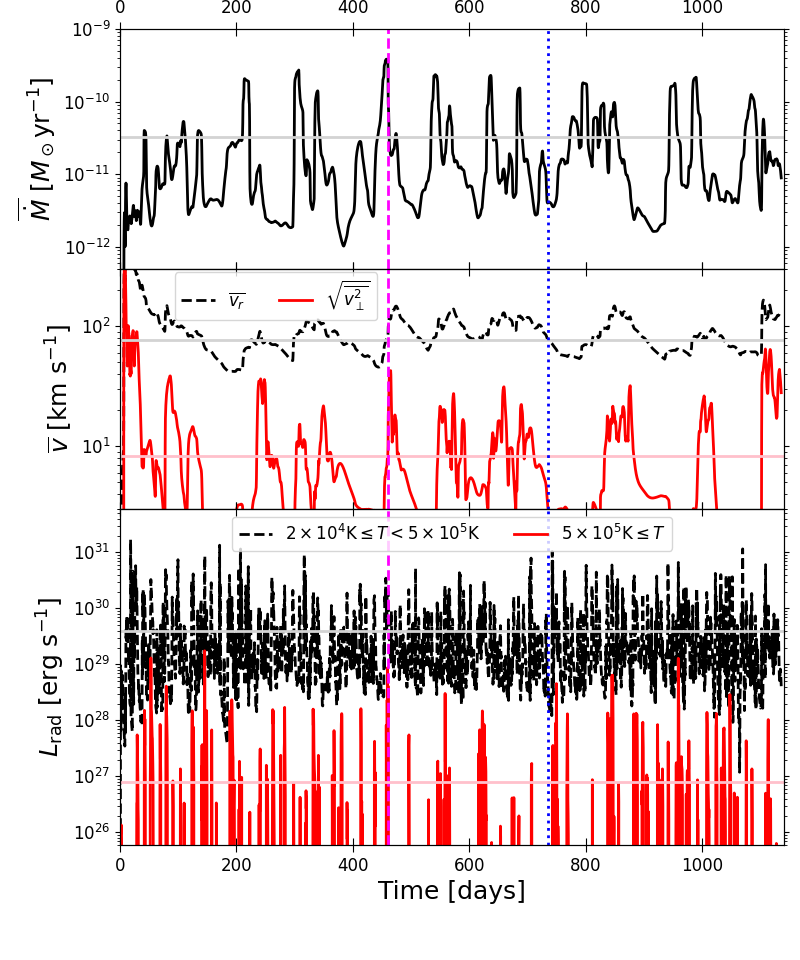}
  \end{center}
  \caption{Time evolution of the mass-loss rate (top panel), 
    {the radial (black dashed line) and rms transverse (red solid line) velocities (middle panel)} and the radiation luminosities (bottom panel) emitted from the gas with $2\times 10^{4}\mathrm{[K]}\le T<5\times 10^{5}$[K] (black dashed line) and  $5\times 10^5\mathrm{[K]}\le T$ (red solid line), respectively, of the nonideal MHD simulation for $\alpha$ Boo. 
    {$\dot{M}$ and $v$ are spatially averaged between $r=28R_{\star}$ and $31R_{\star}$ (see text and equations \ref{eq:vrave} and \ref{eq:vppave}).} The horizontal lines are the time-averaged values of these quantities between $t=113.5$ [days] and $1135$ [days]. The pink and blue vertical lines are respectively active and inactive phases presented in Figure \ref{fig:tav-vs-snp}.
    {Alt text: Three panels from top to bottom respectively present one line, two line and two line graphs. }
  \label{fig:tevol}}
\end{figure}

We examine the time evolution of the mass-loss rate, equation (\ref{eq:masslossrate}), the radial and 
{rms} transverse velocities, and the radiative luminosities from warm and hot atmospheric gas in the nonideal MHD case (Figure \ref{fig:tevol}). Here, $\overline{\dot{M}}$ is the mass-loss rate spatially averaged between $r=28R_\star$ and $31R_\star$ just inside the outer boundary. The velocities are evaluated with the density- and geometrical factor-weighted average, 
\begin{equation}
  \overline{v_r} = \frac{\int dr r^2 f \rho v_r}{\int dr r^2 f \rho}
  \label{eq:vrave}
\end{equation}
and
\begin{equation}
  \sqrt{\overline{v_\perp^2}} = \sqrt{\frac{\int dr r^2 f \rho (v_{\perp,1}^2 + v_{\perp,2}^2)}{\int dr r^2 f \rho}},
  \label{eq:vppave}
\end{equation}
where the integral is taken over the same region as for $\dot{M}$ between $28R_\star$ and $31R_{\star}$.
We assume spherical symmetry when the radiation from the gas 
{at temperatures} between $T_1$ and $T_2$
\begin{equation}
  L_\mathrm{rad}(T_1<T<T_2) = 4\pi \int_{T_1<T<T_2}dr r^2 Q_\mathrm{rad}
  \label{eq:Lrad}
\end{equation}
is calculated. In other words, the radiative flux from magnetically closed regions, which we do not consider in our simulations,  is assumed to be the same as that from open field regions at a given $r$.
As closed loops are supposed to emit larger radiative flux in general than open field regions, this assumption probably underestimates the radiation from the low-altitude atmosphere where the closed structure occupies a nonnegligible volumetric fraction. While we should keep this in mind, we can study characteristic time-variable properties of the radiations at least in a qualitative manner. 

The top panel of Figure \ref{fig:tevol} illustrates that $\overline{\dot{M}}$ greatly varies with time more than two orders of magnitude from $10^{-12}M_{\odot}$yr$^{-1}$ to $4\times 10^{-10}M_\odot$yr$^{-1}$. 
{Both radial and transverse velocities in the middle panel also show large temporal variations. Interestingly, the timings of high $\overline{v_r}$ and $\sqrt{\overline{v_\perp^2}}$ are mostly matched with the periods of large $\overline{\dot{M}}$. We discuss the reason for the correlation between the mass-loss rate and the velocities in Section \ref{sec:rprof}.}

The radiative luminosity 
{of} the warm gas at $2\times 10^4 \,\, \mathrm{K} \le T < 5\times 10^5 \,\, \mathrm{K}$, presented in the bottom panel, also exhibits large time-variabilities over three orders of magnitude. The radiation from the hot gas at $T\ge 5\times 10^5\mathrm{K}$ appears only occasionally; it is present only 16.1\% of the duration  $113.5\; \mathrm{[days]} \le t \le 1135\; \mathrm{[days]}$ used for the time average in Figure \ref{fig:Non-ideal} (see also Section \ref{sec:numsetup})
(see Section \ref{sec:Xobs} for comparison with X-ray observation). 


\subsubsection{Radial Profiles}
\label{sec:rprof}
\begin{figure}
  \begin{center} 
    \includegraphics[height=14.4cm]{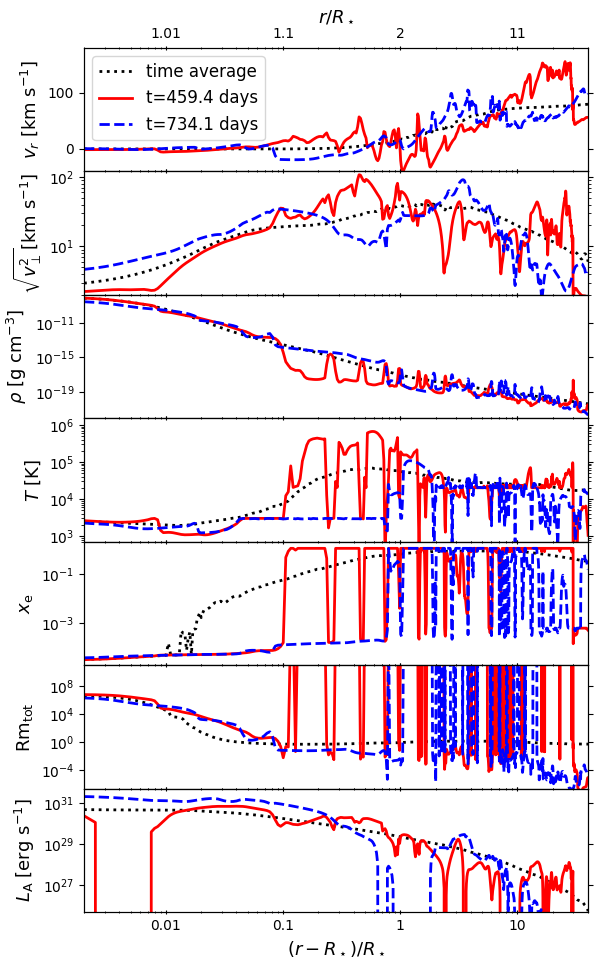}
  \end{center}
  \caption{Comparison of snapshot radial distributions at $t=459.4$ days in an active phase (red solid) and at $t=734.1$ days in an inactive phase (blue dashed) with the time-averaged distribution (black dotted). These timings for the snapshots are also marked in Figure \ref{fig:tevol}. The 
    {seven} panels show, from top to bottom, the radial velocities, the rms transverse velocities, the densities, the temperatures, the ionization degrees, the magnetic Reynolds number for $\eta_\mathrm{tot}$, and the Alfv\'enic Poynting luminosities, respectively. 
    Animation is available as {\it supplementary move} and at \url{https://ea.c.u-tokyo.ac.jp/astro/Members/stakeru/research/movie/index.html}, where different vertical ranges are employed for $v_r$ (top panel) and $\sqrt{v_{\perp}^2}$ (second panel). 
    {Alt text Seven panels from top to bottom respectively present three line graphs. }
  \label{fig:tav-vs-snp}}
\end{figure}

Figure \ref{fig:tav-vs-snp} displays snapshot atmospheric structures at active (red solid) and inactive (blue dashed) phases, in comparison with the time-averaged profile (black dotted), where we selected the (in)active snapshot from a typical period when the radiation from the hot gas at $\ge 5 \times 10^5$~K is present (absent) as  shown in the bottom panel of Figure \ref{fig:tevol} (see also the movie of Figure \ref{fig:tav-vs-snp}). One can clearly see multiple high-temperature $\sim 10^6$~K (fourth panel) and low-density (third panel) hot bubbles in $1.1R_{\star}<r \lesssim 2R_{\star}$ of the active phase. These hot bubbles are adjacent to cool gas with $T\sim$ a few $10^3$~K, which is indicative of the coexistence of fully ionized coronal plasma and molecules \citep{Tsuji2009}. The drastic change in the temperature is due to the thermally unstable temperature range, $T\sim 10^5$~K, in the radiative cooling function \citep{Suzuki2007,Suzuki2018,Washinoue2023}.
A characteristic property of these hot bubbles is that they are supported by the magnetic pressure, $B_\perp^2/8\pi$, associated with Alfv\'{e}nic waves in addition to the gas pressure (see \citet{Suzuki2007} for the detailed formation mechanism of such magnetized hot bubbles).
The second innermost hot bubble around $r\approx 1.4R_{\star}$ triggers radial expansion with $v_r\gtrsim 50$ km s$^{-1}$ (top panel). The rms transverse velocity, $\sqrt{v_{\perp}^2} = \sqrt{v_{\perp,1}^2 + v_{\perp,2}^2}$, also takes large values $\gtrsim 100$ km s$^{-1}$, which are much greater than the time-averaged value of $\approx 30$ km s$^{-1}$ 
at the same radius (second panel).

The transient 
magnetized bubbles 
are the origin of the large time-variability in $\dot{M}$ (Figure \ref{fig:tevol}).
One can 
recognize an evident shock structure consisting of a high-density region with a clear velocity jump at $r\approx 30 R_{\star}$ at the active phase (red solid lines in Figure \ref{fig:tav-vs-snp}). This originates from multiple magnetized hot bubbles formed in the inner region $r \lesssim 2R_{\star}$ at $\approx 30-50$ days earlier. These bubbles sweep up the lower-radial velocity gas in front to form the shock (see the animation version of Figure \ref{fig:tav-vs-snp}), which causes the enhancement in $\dot{M}$ and $v_r$ at $t\approx 460$ days in Figure \ref{fig:tevol} (vertical pink dashed line). In other words, the magnetic activity, accompanied by hot bubbles in the inner region, leads to high-velocity and dense winds in the outer region several tens of days later.

{In contrast to the large temperature variation in $r\lesssim 3R_{\star}$, the temperature mostly remains below $\sim \! 2 \times 10^4$~K in $r\gtrsim 3R_{\star}$ except for the shock front at $r\approx 30 R_{\star}$. This is because Lyman cooling inhibits the gas from heating up beyond $\sim 2\times 10^4$~K .}

The temperature in the inactive period stays low 
in the inner region of $r\lesssim 2R_{\star}$.  Accordingly, the density is higher than the time-averaged value (second panel of Figure \ref{fig:tav-vs-snp}) and the gas 
{in $1.08R_{\star}\le r\le 1.5R_{\star}$} 
falls back (i.e., $v_r <0$) to the surface (top panel). 
Similar downflows with 
$v_r$ down to $-50$~km s$^{-1}$ are frequently observed in inactive phases (See the animation of Figure \ref{fig:tav-vs-snp}).
We note that the plateau at $T\approx 3000$ K in $r\le 1.7 R_{\star}$ corresponds to the cut-off temperature $T_\mathrm{cut}=3010$ K of the radiative cooling, below which the cooling is artificially switched off (Section \ref{sec:MHDeqs}).
However, we also note that the temperature can drop below $T_\mathrm{cut}$ by adiabatic expansion. There are multiple low-temperature and high-density regions in both active and inactive phases; the temperature of some of such dense blobs is $\lesssim 1000$K. These cold regions are sandwiched by warm or hot bubbles, 
leading to the multi-temperature layers 
{ranging from} $10^3$K to $10^5$--$10^6$K 
coexisting in the atmosphere \citep[see][for observations]{Harper2013,Ohnaka2018}. 

Reflecting the spatially inhomogeneous temperature distributions, fully ionized regions with $x_e\approx 1$ and weakly ionized regions with $x_e\ll 1$ coexist in both active and inactive phases as shown in the fifth panel of Figure \ref{fig:tav-vs-snp}. The magnetic Reynolds numbers, which are proportional to $x_e$ in the weakly ionized plasma (see equations \ref{eq:etaO} -- \ref{eq:Rm}), also largely vary with $r$. In the fully ionized region, $\eta_\mathrm{tot}\Rightarrow 0$, and hence, Rm$_\mathrm{tot}\Rightarrow \infty$, which is outside the displayed range of the sixth panel. The large 
{spatial and temporal} variation in Rm$_\mathrm{tot}$ 
causes the 
{difference} between the 
{time-averages of} $\langle \mathrm{Rm_{tot}} \rangle$ calculated using the arithmetic and harmonic means as shown in the 
fourth panel of Figure \ref{fig:Non-ideal}. 

The bottom panel of Figure \ref{fig:tav-vs-snp} compares the snapshot and time-averaged Alfv\'enic Poynting luminosities. While the snapshot profiles of $L_\mathrm{A}$ in both active and inactive phases roughly follow the trend of the time averaged $\langle L_\mathrm{A} \rangle$, both snapshots take negative values in some places, where $L_\mathrm{A}$ falls out of the displayed range; in these regions, 
incoming Alfv\'enic waves dominate outgoing ones as a result of the continual excitation of reflected waves.
The density fluctuations associated with the large temperature variations (top and second panels) act as ``density mirrors'' in the reflection of Alfv\'enic waves because they cause changes in $v_\mathrm{A}$ with scales smaller than the wavelength \citep{Suzuki2006JGRA,Shoda2018b}.

\subsubsection{Observational Implications}
\label{sec:obsimp}
We would like to discuss how the large temporal variations simulated within a single flux tube are observed. The stellar surface is considered to be covered by a number of flux tubes (see Section \ref{sec:numsetup}), each of which behaves somewhat independently. Therefore, from an observational perspective, we expect 
these temporal variations 
to also manifest themselves  as spatial inhomogeneities 
in the horizontal direction over the surface. 

The rms transverse velocity shown in the second panel of Figure \ref{fig:tav-vs-snp} can be compared with the turbulent velocities 
{derived} from observations. The time-averaged $\sqrt{v_\perp^2}$ reaches a maximum value of $\approx 40$ km s$^{-1}$ at $r=(2-3)R_\star$, which is slightly 
{higher} than the turbulent velocities of $20-30$ km s$^{-1}$ 
{derived by \citet{Harper2022} based on} the 
HST UV 
line profile analysis.
 This is partly because the time-averaged density in the simulation is slightly lower than the density adopted in \citet{Harper2022}. Additionally, the simulated $\sqrt{v_\perp^2}$ is greatly variable from $\lesssim 10$km s$^{-1}$ to 100 km s$^{-1}$. In reality, there exist simultaneously multiple flux tubes with different $\sqrt{v_\perp^2}$ that behave independently. This inhomogeneous effect needs to be taken into account when estimating turbulent velocities from observations.

The presence of time-variable multi-temperature layers requires a careful treatment when we compare the simulation results with analyses of observational data, which usually assume 
spherically symmetric atmospheric structures. For example, the 
mass-loss rate for $\alpha$ Boo is observationally determined to reproduce the wind-scattered UV emission line profiles 
by assuming a spherical 
wind for $r > 1.2R_{\star}$ \citep{Harper2022}.  However, if the atmosphere is spatially and temporally inhomogeneous, as shown in Figure \ref{fig:tav-vs-snp}, the mass-loss rate estimated from the wind-scattered chromospheric line profiles may underestimate the actual mass-loss rate because the contribution from the gas in the other temperature ranges is omitted.

On the one hand, the mass fraction of high-temperature coronal gas in the total wind material is minimal as the density is low, and therefore, the contribution of the high-temperature component to the mass loss is negligible. On the other hand, low-temperature gas could occupy a significant portion of the total mass loss. The mass fraction of the cool-component gas 
at $T\le 10^4$~K 
in $1.2 R_{\star}\le r \le 10 R_{\star}$ is 28\% and 62\% at the active and inactive times presented in Figure \ref{fig:tav-vs-snp}, respectively, where the radial integration range is adopted 
{for a comparison} with the wind model 
{of} \citet{Harper2022}. 
The mass fraction of the cool component averaged over time between $0.1t_\mathrm{sim}$ and $t_\mathrm{sim}$ is 45\%;  
the mass-loss rate estimated from the 
chromospheric material alone is 55\% of the total mass-loss rate.

While our model is still incomplete particularly for the treatment of cool gas because of the ignorance of the radiation cooling below $T=T_\mathrm{cut}$ and the latent heat (Sections \ref{sec:MHDeqs} and \ref{sec:ion}), it might be useful to see how  
the cool, dense blobs found in the simulation compare with
the observations of 
{the extended molecular outer} atmosphere -- the so-called MOLsphere \citep{Tsuji2009,Ohnaka2018}. 
\citet{Ohnaka2018} spatially resolved the MOLsphere of $\alpha$ Boo in the 2.3~$\mu$m CO lines with VLTI/AMBER and observationally determined its physical properties. 
{The CO layers extend to $\sim$2.6~$R_{\star}$ with temperatures of 1600--1800~K, somewhat cooler than predicted by the model even at the inactive phases.}
They derived densities of 
{$1.5\times 10^{-11}$ and} $3.7\times 10^{-14}$~g cm$^{-3}$ 
{at 1.04 and 2.6~$R_{\star}$, respectively}.
The densities of 
$\sim 10^{-12}-10^{-11}$ and $\sim 10^{-17}-10^{-16}$~g~cm$^{-3}$ of the cool dense blobs in our simulation at the same radii (Figure \ref{fig:tav-vs-snp}) are lower than the observed values. 
On the one hand, the discrepancy indicates that a more sophisticated prescription for the cool component is required in our simulations.
Additionally, multi-dimensional effects, which are not considered in our 1D treatment, are supposed to be also important because compression in the transverse direction
could create locally dense regions. 
On the other hand, from an observational point of view, 
\citet{Ohnaka2018} used a simple, spherical, two-layer model to interpret the 
spatially resolved observational data, while their data show signatures of inhomogeneities in the MOLsphere. An analysis of the interferometric data with more sophisticated models is also necessary to relate the presence of the MOLsphere to our simulation.

\subsubsection{X-ray}
\label{sec:Xobs}

\citet{Ayres2003} and \citet[][]{Ayres2018} detected weak X-ray emissions from $\alpha$ Boo with 
luminosities of $1.5\times 10^{25}$ 
(tentative $3\sigma$ detection)  and $3\times 10^{25}$~erg~s$^{-1}$ in 2002 and 2018, respectively,
in the energy range of 0.2 - 2 keV with {\it Chandra} High Resolution Camera. 
\citet{Schmitt2024} also reported detection of X-ray emission in 2021 with {\it XMM-Newton} with a luminosity of $2\times 10^{25}$~erg~s$^{-1}$ in the same energy range.

Our fiducial run for $\alpha$ Boo also exhibits time-dependent high energy radiation in the EUV and soft X-ray range emitted from 
{the} transiently formed hot coronal gas (Figure \ref{fig:tevol}). To directly compare to these {\it Chandra} observations, we analyze the hot plasma 
with $T>2\times 10^6$~K, which roughly corresponds to $>0.2$ keV, in the simulation data. If we use equation (\ref{eq:Lrad}) by assuming spherical symmetry, the X-ray luminosity transiently becomes $10^{27}-10^{28}$~erg s$^{-1}$ at its maximum level.

However, as explained in Section \ref{sec:numsetup}, $\approx 2800$ open magnetic flux tubes cover the stellar surface, and these flux tubes behave 
  more or less independently and emit X-ray transiently in a stochastic manner. Hence, the X-ray luminosity from each tube is 1/2800 of this value, giving $\sim 10^{24}$~erg s$^{-1}$ at the maximum.  The duration of such strong X-ray emissions from a single tube is 
$\lesssim 0.1\%$ of the total simulation time. 
Therefore, on average, $\lesssim$ a few flux tubes are actively emitting X-rays with $10^{24}-10^{25}$~erg s$^{-1}$ in the $>0.2$ keV range; if $\gtrsim 10$ flux tubes are active at the same time by stochasticity, the observed luminosities could be reproduced in such instances.

There are several caveats to be addressed in this argument.
{First and foremost, the contribution from closed magnetic loops need to be considered, because they are expected to emit stronger X-rays} (Section \ref{sec:timevariation}).
{Additionally, from an observational point of view, it should be noted that the magnetic activity of $\alpha$ Boo is variable with a period of $\ge$14 years \citep{Brown2008}. Therefore, it is possible that the X-rays detected by \citet{Ayres2003}, \citet{Ayres2018}, and \citet{Schmitt2024}} are higher than at other epochs.

\subsection{Dependence on $B$ }
\label{sec:depB}
\begin{figure}
  \begin{center}
    \includegraphics[height=6cm]{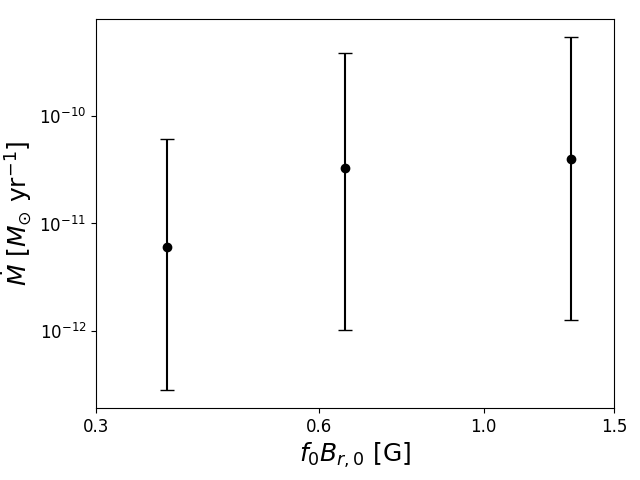}
  \end{center}
  \caption{Dependence of the mass-loss rate on the average magnetic field strength, $f_0B_{r,0}$. The dots indicate the time-averaged values between $t=113.5$ [min] and 1135 [min] and the 
    vertical bars are determined by the maximum and minimum values during this period. 
    {Alt text: Three data points with error bars.}
  \label{fig:MdotBdep}}
\end{figure}

We investigate the effects of the magnetic field on 
the structure of RGB stellar winds.
We vary the spatially averaged field strength, $f_0 B_{r,0}$, by changing the filling factor (equation \ref{eq:fillfac}) while keeping $B_{r,0}$ at the equipartition value at the photosphere (equation \ref{eq:p0}). Figure \ref{fig:MdotBdep} shows that the three cases with different $f_0 B_{r,0}$ exhibit 
large time variabilities over two orders of magnitude.
The time-averaged $\langle \dot{M}\rangle$ resulting from the case with a weaker magnetic field of $f_0 B_{r,0}=0.37$~G is about 1/5 of that in the fiducial case with $f_0 B_{r,0}=0.65$~G (Table \ref{tab:runsBoo}). While the maximum $\dot{M}$ of the weak $f_0B_{r,0}$ case is larger than the average $\langle\dot{M}\rangle$ of the fiducial case because of the large temporal variation,  the dependence of $\langle \dot{M}\rangle$ on $f_0 B_{r,0}$ is very sensitive.
On the other hand, $\langle \dot{M}\rangle$ in the stronger $f_0 B_{r,0}$ ($=1.31$ G) case is only 20\% larger than in the fiducial case.

\begin{figure}
  \begin{center}
    \includegraphics[height=11cm]{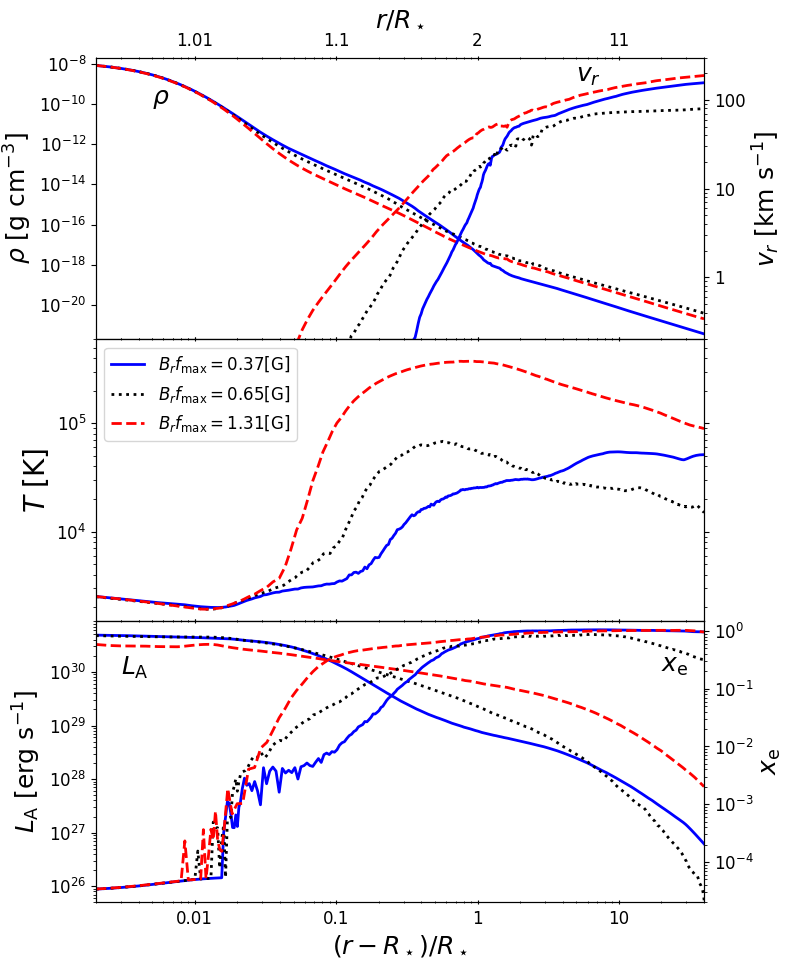}
  \end{center}
  \caption{Comparison of the time-averaged radial wind profiles of the cases with different $f_0 B_{r,0}= 0.37$ G (blue solid), 0.65 G (black dotted), and 1.31 G (red dashed). The physical quantities shown in the three panels are the same as in the top three panels of Figure \ref{fig:Non-ideal}.
  {Alt Text Three panels from top to bottom respectively present six line, three line, and six line graphs.}
  \label{fig:str_B}}
\end{figure}

In order to understand this 
{nonlinear} dependence on $f_0 B_{r,0}$ 
{with saturation}, we inspect the radial atmospheric structures in Figure \ref{fig:str_B}. Let us begin with the Alfv\'enic Poynting flux presented in the bottom panel (left axis). The weak field case (blue solid) shows a rapid decrease in $L_\mathrm{A}$ around $r\approx 1.1 R_{\star}$ because the Alfv\'enic waves take more time to travel through this region owing to the lower $v_\mathrm{A}(\propto B_r\propto f_0B_{r,0})$ (equation \ref{eq:divB}) and suffer ambipolar diffusion more severely.
In the static condition, $v_r\ll v_\mathrm{A}$,  the Alfv\'enic Poynting flux in equation (\ref{eq:LA}) can be approximated as 
{$L_\mathrm{A}\propto B_\perp^2v_\mathrm{A}/(4\pi)$}
, where we used the relation, 
{$v_\perp = -B_\perp/\sqrt{4\pi\rho}$}, 
for the outgoing Alfv\'en wave (equation \ref{eq:zpp}). Therefore, the rapid decrease in $L_\mathrm{A}$ enhances the outward force due to magnetic pressure (
{$=B_{\perp}^2/(8\pi)$}) gradient, which lifts up the gas. Consequently,  the density in $r\lesssim 2R_\star$ of this case is higher than in the cases with larger $f_0 B_{r,0}$ (top panel).

However, the temperature in the same region is lower in the weak field case than in the other cases (middle panel) although more energy is supplied to the gas by the dissipation of $L_\mathrm{A}$. This is because the radiative cooling, which is enhanced by the increase in $\rho$ (equations \ref{eq:Qrad_hT} and \ref{eq:Qrad_lT}), 
offsets the heating 
{due to} the dissipation of $L_\mathrm{A}$.
The lower-temperature condition 
also 
{leads to the} lower ionization degree in $r\lesssim 2R_\star$ (right axis of bottom panel), which further promotes the ambipolar diffusion. 
As a result, the two cases with $f_0 B_{r,0} = 0.37$ G and 0.65 G give a significant difference in $L_\mathrm{A}$ that reaches the higher-altitude, $r\gtrsim 2R_{\star}$ (bottom panel), where the stellar wind is accelerated (right axis of top panel). This difference directly leads to the difference in the mass-loss rate as shown in Figure \ref{fig:MdotBdep} and Table \ref{tab:runsBoo}. In summary, the sensitive dependence of $\dot{M}$ on $f_0 B_{r,0}$ can be interpreted by the positive feedback through the ambipolar diffusion of Alfv\'enic waves in the lower-atmosphere, $r\lesssim 2R_{\star}$. 

Let us turn to the stronger field case with $f_0 B_{r,0}=1.31$ G (red dashed lines in Figure \ref{fig:str_B}). The Alfv\'enic Poynting flux $L_\mathrm{A}$ in the 
{innermost} region
{ $r\le 1.08R_{\star}$} 
is lower than 
{that of} the weaker field cases (bottom panel). This is because a larger fraction of the outgoing Alfv\'enic waves is reflected downward to the surface; here we note that $L_\mathrm{A}$ is the net outgoing Poynting flux, which corresponds to the outgoing component minus the incoming component \citep{Suzuki2013}. 
{The reflection of the Alfv\'enic waves is promoted as follows.} 
{The} wavelength of an Alfv\'en wave is proportional to the local $v_\mathrm{A}$
{, which is in turn } $\propto B_r$. 
{This means that} the wavelength for a given frequency is 
{longer} with a larger $f_0 B_{r,0}$. Therefore, it easily becomes 
{longer} than the radial variation scale of $v_\mathrm{A}$, which promotes the reflection \citep{Suzuki2006JGRA}. 

On the other hand, the decrease in $L_\mathrm{A}$ 
{in the strong field} case is slower at $r \gtrsim 1.1 R_{\star}$ 
than in the other cases. The reason is the same as 
{in} the weak field case mentioned above; the Alfv\'enic waves propagate more rapidly through the lower atmosphere because of the higher $v_\mathrm{A}$ 
{without significant damping due to the} ambipolar diffusion. As a result, the uplift of the gas is suppressed, and the lower-density gas (top panel) can be heated up to 
{higher temperatures} (middle panel) than in the other cases.
The inhibition of the wave damping results in a reduction in the heating of the gas, which suppresses the mass loading to the wind. The density in the outer region 
{in} this case is slightly lower than that 
{in} the fiducial case. Therefore, the mass-loss rates are almost comparable 
{each other}, although the velocity 
{in} the stronger field case is considerably higher (top panel) 
{because of} the efficient acceleration of the lower-density gas.

The radiative luminosities from the hot and warm gas also show  
dependences on $f_0B_{r,0}$ 
{similar to} the mass-loss rate (Table \ref{tab:runsBoo}) since both $L_\mathrm{rad}$ and $\dot{M}$ 
show positive 
{correlations with} the density. In a quantitative sense, 
the time-averaged $\langle L_\mathrm{rad}\rangle$ from the highest-temperature component at $T\ge 5\times 10^5$~K shows a steeper dependence on $f_0B_{r,0}$ than that of the 
{warm} component at $T = 2 \times 10^4 - 5\times 10^5$~K. 
This is because the magnetized hot bubbles are formed one after another, and
{the $L_\mathrm{rad}$ from the hot gas is significant for a larger fraction of {the simulation} time for a larger $f_0B_{r,0}$.}
For example, in the case with $f_0B_{r,0}=1.31$~G,  
the hot bubbles 
at $T>5\times 10^5$ K are present about $2/3$ of the simulation time, which is much larger than 16.1\% in the fiducial case with $f_0B_{r,0}=0.65$~G. 

\subsection{Dependence on Metallicity}
\begin{figure}
  \begin{center}
    \includegraphics[height=6cm]{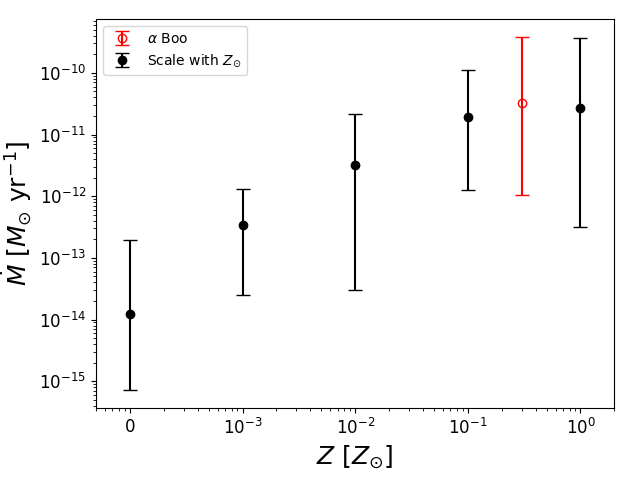}
  \end{center}
  \caption{Same as Figure \ref{fig:Mdotsum} but for the {metallicity} dependence 
    The red open symbol indicates the result of the 
    {fiducial} run for $\alpha$ Boo. {Alt text: Six data points with vertical error bars.}
  \label{fig:Mdotzdep}}
\end{figure}

\begin{figure}
  \begin{center}
    \includegraphics[height=11cm]{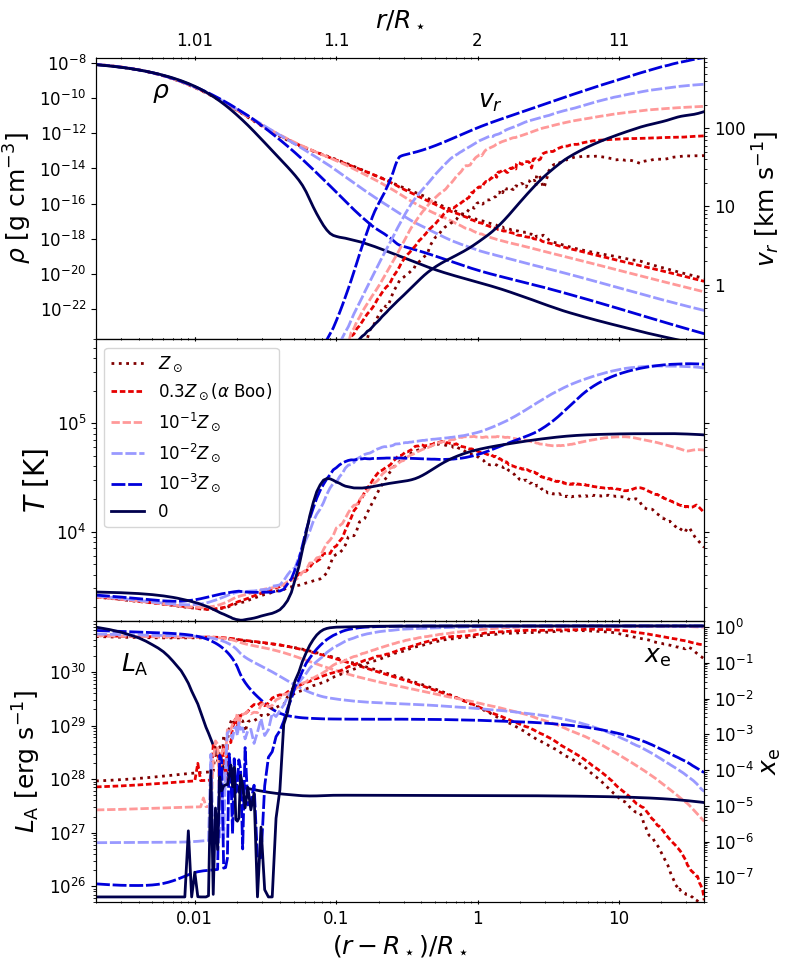}
  \end{center}
  \caption{Same as Figure \ref{fig:str_B} (time-averaged quantities) but for the cases with different metallicities. Shorter dashed, 
    {darker red} lines 
    correspond to {the} higher-metallicity cases. 
    Three panels from top to bottom respectively present twelve line, six line, and twelve line figures.
  \label{fig:str_metal}}
\end{figure}

{The metallicity} is an important parameter that controls the physical properties of RGB winds with the nonideal MHD effect because the electrons in the low-temperature atmosphere are mainly supplied from the ionization of heavy elements.
We conduct simulations by varying the metallicity, as shown in Table \ref{tab:runsBoo}, while keeping the other stellar parameters fixed. We should note that, in reality, stellar parameters such as the photospheric density and the effective temperature are also altered because the opacity is substantially 
{governed} by the metallicity \citep[see, e.g.,][for solar-type stars]{Suzuki2018}.  Furthermore, the evolutionary track itself is also dependent on 
{the} metallicity. In order to avoid these complexities and isolate the {metallicity} effect, 
we change only the metallicity that determines the ionization degree and radiative cooling.

Figure \ref{fig:Mdotzdep} demonstrates the positive dependence of the mass-loss rate on {the} metallicity in the range of $Z\le 0.1Z_\odot$. The bottom panel of Figure \ref{fig:str_metal} illustrates that the ionization degree (right axis) is lower and the Alfv\'enic Poynting flux decreases more rapidly in {the} lower $Z$ cases.
Consequently, the density in the wind region is lower (left axis of top panel), leading to smaller $\dot{M}$. The lower-density gas is accelerated to higher velocities (right axis of top panel) and heated up to higher temperatures (middle panel).

The velocity and temperature of the zero-metal case (black solid lines in Figure \ref{fig:str_metal}) 
{deviate} from the abovementioned 
dependence on the metallicity; $v_r$ and $T$ are not as high as those in the low-metallicity cases with $Z=10^{-3}Z_{\odot}$ and $10^{-2}Z_{\odot}$.  This is because $L_\mathrm{A}$ drastically drops by three orders of magnitude 
{at} $r\le 1.02 R_{\star}$
{, where the ionization degree is very low, $x_e\approx 3\times 10^{-8}$, which results only from the ionization of hydrogen by the interstellar radiation field} (Section \ref{sec:ion}). Hence, the stellar wind is extremely weak, yielding the slow rise in $v_r$ and lower $T$ than in the low $Z$ cases. 

{On} the metal-rich side, the mass-loss rate is saturated for $Z>0.1Z_{\odot}$; the time-averaged $\langle\dot{M}\rangle$ for $\alpha$ Boo is slightly larger than $\langle\dot{M}\rangle$ of the solar metallicity case (Table \ref{tab:runsBoo}), whereas we note that, although $\alpha$ Boo is labeled as $Z=0.3Z_{\odot}$ from the iron abundance, some heavy elements, such as O and Mg, are as abundant as in the Sun (Table \ref{tab:metal}).
The main reason of this saturation in $\dot{M}$ is the increase in the radiative loss for the metal-rich condition. The density in the solar metallicity case is {the} highest 
{above $r=1.1R_{\star}$} (top panel of Figure \ref{fig:str_metal}) among the presented cases
, because the sufficient $L_\mathrm{A}$ that reaches the wind region leads to efficient mass loading. This is 
{because the sufficient ionization degree in the lower atmosphere 
  {suppresses} ambipolar diffusion} (bottom panel). On the other hand, the higher density also promotes {the} radiation cooling as shown in Table \ref{tab:runsBoo} following equations \ref{eq:Qrad_hT} and \ref{eq:Qrad_lT}, 
reducing $L_\mathrm{A}$ in the outer region (bottom panel of Figure \ref{fig:str_metal}). The final wind velocity in the solar metallicity case is $\langle v_r\rangle\approx 40-50$ km s$^{-1}$, which is considerably lower than $\langle v_r\rangle\approx 70-80$ km s$^{-1}$ obtained from the fiducial case for $\alpha$ Boo (top panel). As a result, $\langle \dot{M}\rangle$ in the former case is moderately lower than in the latter case even though the density is slightly higher.

\section{Aldebaran}
\label{sec:Tau}
\begin{figure}
  \begin{center}
    \includegraphics[height=11cm]{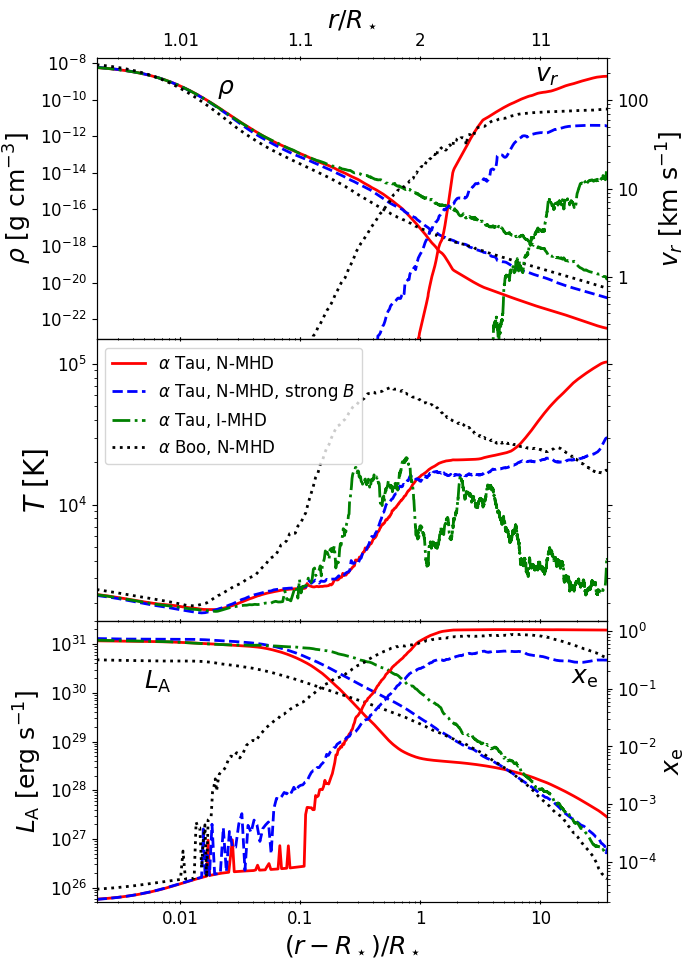}
  \end{center}
  \caption{Same as Figure \ref{fig:str_B} (time-averaged quantities) but for the comparison of $\alpha$ Tau (red solid and blue dashed lines for the nonideal MHD runs with $f_0 B_{r,0}=0.25$ G and 0.50 G, respectively, and green dot-dashed lines for the ideal MHD run) with $\alpha$ Boo (non-ideal MHD run; black dashed lines). 
    Note that $x_e$ in the ideal MHD case is not shown. 
    Three panel from top to bottom respectively present eight line, four line, and seven line figures.
  \label{fig:str_aTau}}
\end{figure}

We conduct MHD simulations of three different cases for $\alpha$ Tau (Table \ref{tab:runsTau}). First, we show results for the nonideal and ideal simulations with observationally compatible magnetic field strength, $f_0B_{r,0}=0.25$ G in Section \ref{sec:Tau_weak}. Additionally, we perform a nonideal MHD run with {a} stronger field {with} $f_0 B_{r,0}=0.50$ G in Section \ref{sec:Tau_strong}. 

\subsection{Results with {a} Fiducial Magnetic Field of $f_0B_{r,0}=0.25$~G}
\label{sec:Tau_weak}
The mass-loss rates in both ideal and nonideal cases are considerably smaller than in the counterpart runs for $\alpha$ Boo (Figure \ref{fig:Mdotsum} and Table \ref{tab:runsTau}) because the magnetic field with $f_0B_{r,0}=0.25$ G is weaker than the 0.65 G for $\alpha$ Tau. 
The time-averaged $\langle\dot{M}\rangle=1.5\times 10^{-12}M_{\odot}$ yr$^{-1}$ 
in the nonideal MHD case is more than one order of magnitude lower than that obtained 
{in} $\alpha$ Boo (Table \ref{tab:runsBoo}). This value is also lower than the observationally estimated mass-loss rate of 
$(1.0-1.6)\times 10^{-11}M_{\odot}$ yr$^{-1}$ \citep{Robinson1998,Wood2007}, whereas the snapshot $\dot{M}$ transiently becomes $\dot{M}_\mathrm{max} = 1.4\times 10^{-11}M_\odot$ yr$^{-1}$. 
The reason of the weak wind in this case is the severe ambipolar diffusion of Alfv\'enic waves. 

In Figure \ref{fig:str_aTau} we present the time-averaged wind profiles of the 
{the three cases} for $\alpha$ Tau in comparison with those of the {fiducial} nonideal MHD run for $\alpha$ Boo (black dashed).  The ionization degree in the nonideal MHD run with $f_0B_{r,0}=0.25$ G (red solid) 
remains very low, $x_e\lesssim 10^{-4}$, in a broad 
region of $\alpha$ Tau (right axis of bottom panel), mainly because the magnetic field is weaker (see Section \ref{sec:depB} for the $B$ dependence). In addition, the radiative ionization is also slightly suppressed because of the lower $T_\mathrm{eff}$ {compared to $\alpha$~Boo} (see Table \ref{tab:stprm} and equation \ref{eq:Saha}).
As a result, the Alfv\'enic Poynting flux {$L_\mathrm{A}$} is greatly reduced (left axis of bottom panel) by the ambipolar diffusion 
{in marked contrast to} the ideal MHD case (green dot-dashed line in bottom panel). With the drop in $L_\mathrm{A}$, the density also rapidly decreases (left axis of top panel), 
{resulting in} the weak stellar wind {with a significantly lower mass-loss rate}. On the other hand, the 
{terminal wind} velocity and temperature (top and middle panels) are higher because the lower-density gas can be more easily accelerated and heated. 


\subsection{Strong Magnetic Fields?}
\label{sec:Tau_strong}

\citet{Auriere2015} show that the longitudinal magnetic field strength in Aldebaran varied at three epochs: non-detection at the first epoch, definitive detection at the second epoch, and definitive detection with a polarity reversal at the third epoch (see also Section \ref{sec:Bfield}). They also note that while the measured magnetic field strengths are of the sub-G level, another indicator of the magnetic activity derived from the Ca II H and K lines ($S$-index) for $\alpha$ Tau is as large as in the stars with field strengths of a few G.\footnote{Although it is not as extreme as $\alpha$ Tau, a similar tendency is obtained for $\alpha$ Boo in \citet{Auriere2015} as discussed in Section \ref{sec:Bfield}.} \citet{Auriere2015} suggest that $\alpha$ Tau possesses multiple small-scale magnetic regions of opposite polarities, which cancel out in the surface-averaged field strength, resulting in the weak and variable signals.
Therefore, it is possible that the actual field strengths of individual magnetic flux tubes in $\alpha$ Tau are stronger than assumed in our present fiducial setup.

\begin{figure}
  \begin{center} 
    \includegraphics[height=14.4cm]{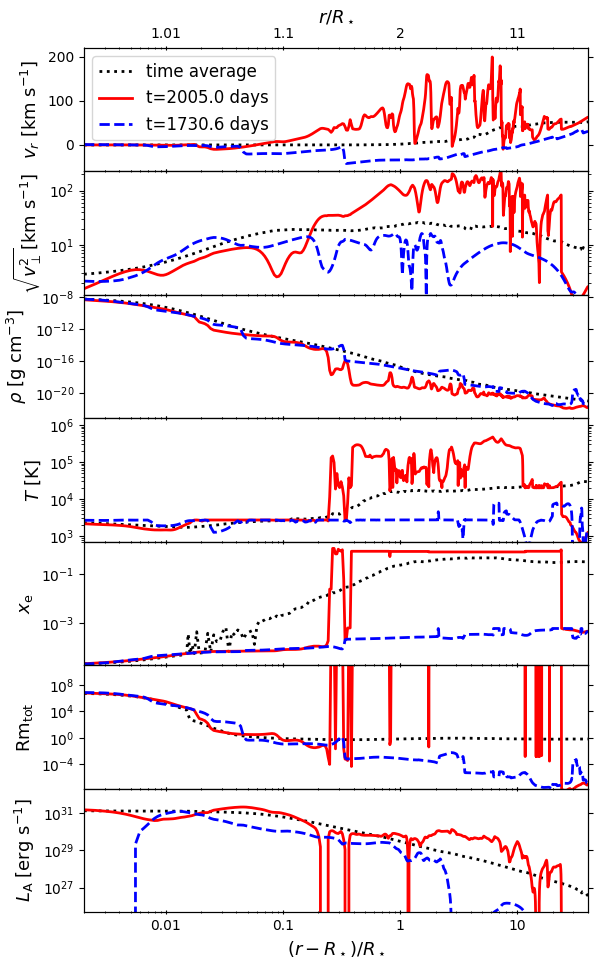} 
  \end{center}
  \caption{Same as Figure \ref{fig:tav-vs-snp} (snapshots) 
  but for 
    {the nonideal MHD run with $f_0B_{r,0}=0.50$ G for} $\alpha$ Tau. The radial distributions at $t=$ 
    {2005.0} days and 
    {1730.6} days 
    represent typical active (red solid) and inactive (blue dashed) periods.
    Animation is available as {\it supplementary move} and at \url{https://ea.c.u-tokyo.ac.jp/astro/Members/stakeru/research/movie/index.html}.
   Sevel panels from top to bottom respective present three line figures.
  \label{fig:tav-vs-snp_aTau}}
\end{figure}

In light of this 
possibility, we also present the results with a stronger magnetic field of $f_0B_{r,0} = 0.5$~G in Figure \ref{fig:str_aTau} (blue dashed lines). In the wind region, the density is enhanced 
compared to the standard case with $f_0B_{r,0}=0.25$~G (red solid line) by nearly two orders of magnitude. By contrast, the velocity is reduced to $\langle v_\mathrm{out}\rangle \approx 50$~km~s$^{-1}$ (Table \ref{tab:runsTau}), which is comparable to the observationally estimated values of  $30-40$~km~s$^{-1}$ \citep{Robinson1998,Wood2016}. 
The time-averaged mass-loss rate increases to $2.0\times 10^{-11}M_{\odot}$ yr$^{-1}$ by more than one order of magnitude because of the enhancement in the density (Table \ref{tab:runsTau}). 
This predicted mass-loss rate is comparable to the observed values {of $(1.0-1.6)\times 10^{-11}M_{\odot}$ yr$^{-1}$} \citep{Robinson1998,Wood2007} and $2.8\times 10^{-11}M_{\odot}$ yr$^{-1}$ \citep{Wood2024}. 
We note that the sensitive dependence of $\langle\dot{M}\rangle$ on $f_0B_{r,0}$ is consistent with the trend obtained for $\alpha$ Boo (Section \ref{sec:depB})

Figure \ref{fig:tav-vs-snp_aTau} demonstrates snapshot wind structures 
{in the strong field case with} $f_0B_{r,0}=0.50$ G. 
One can see multiple hot bubbles with $T\approx 5\times 10^5$~K distributed from the low atmosphere to the wind region ($\gtrsim 10R_{\star}$). 
However, we note that, while such magnetized hot bubbles manifest themselves multiple times during the simulation time, the total time fraction 
with the presence of these hot bubbles is considerably 
smaller than in the fiducial nonideal MHD run for $\alpha$ Boo (see the movie version of Figure \ref{fig:tav-vs-snp_aTau}). 
Tables \ref{tab:runsBoo} and \ref{tab:runsTau} show that the time-averaged luminosity {from the gas at $T>5\times 10^5$~K $\langle L_\mathrm{rad}(5\times 10^5{\rm K}<T)\rangle$ is much lower, $3.1\times 10^{25}$~erg~s$^{-1}$, for $\alpha$~Tau 
than the $7.9\times 10^{26}$~erg~s$^{-1}$ in the case of $\alpha$ Boo.}

In the inactive 
{phase} (blue dashed lines), the temperature stays 
{$\lesssim 10^4$} K in the entire simulation domain, and the complete ionization is not realized even in the outer region (fifth panel). The ambipolar diffusion is still at work there to give small Rm$_\mathrm{tot}<1$ (sixth panel), and Alfv\'enic waves are severely damped (bottom panel). We should mention that the flat temperature distribution in 
{$r\lesssim 10R_{\odot}$} is due to the cut-off of the radiation cooling at $T_\mathrm{cut}=2730$ K (Section \ref{sec:MHDeqs}); more elaborated treatment is required for cooler gas, which will allow 
{us to study the formation of molecules in the outer atmosphere and compare with the observations in a more quantitative manner} \citep{Tsuji2001,Tsuji2008,Ohnaka2013}, as in the case of $\alpha$~Boo discussed in Section~\ref{sec:obsimp}. 

{The} simulations 
with different magnetic field strengths exhibit that the thermal and dynamical properties of the chromosphere are substantially controlled by the magnetic field. 
In reality, multiple flux tubes with different magnetic field strengths are distributed over the stellar surface. The mass-loss and thermal properties are observed as the sum of the contributions from these different flux tubes (see Sections \ref{sec:massloss} and \ref{sec:obsimp}). 



\section{Summary and Discussion}
\label{sec:sum}
By performing nonideal MHD simulations, we studied stellar winds from typical 
non-coronal K giants, $\alpha$ Boo and $\alpha$ Tau. Alfv\'enic waves excited by surface convection are damped by ambipolar diffusion as they propagate through the weakly ionized atmosphere.
In the simulations for $\alpha$ Boo, the time averaged mass-loss rate, $\langle\dot{M}\rangle=3.3\times 10^{-11}M_{\odot}$ yr$^{-1}$, is reduced by the ambipolar diffusion more than one order of magnitude, compared to the ideal MHD case, and is consistent with the observationally derived value by \citet{Harper2022}. The nonideal MHD simulation for $\alpha$ Tau results in smaller $\langle \dot{M} \rangle=1.5\times 10^{-12}M_\odot$ yr$^{-1}$ because 
the 
observed surface-averaged magnetic field strength is weaker and Alfv\'enic waves are more severely affected by ambipolar diffusion in the lower-temperature atmosphere. 

The mass-loss rate is positively dependent on the magnetic field strength with saturation. An increase in the field strength,  
which leads to an increase in the Alfv\'en velocity, makes Alfv\'enic waves travel more rapidly through the weakly ionized region, suppressing the ambipolar diffusion and raising the $\dot{M}$. At the same time, it also promotes the reflection of Alfv\'enic waves, causing the saturation of the $\dot{M}$ for further stronger fields.
Although this tendency is obtained from the simulations with stellar parameters for $\alpha$ Boo, it is 
{directly applicable} to $\alpha$ Tau. 
If we adopt a magnetic field strength twice as strong as the original, the time-averaged mass-loss rate increases by more than one order of magnitude to $\langle\dot{M}\rangle = 2.0\times 10^{-11}M_{\odot}$yr$^{-1}$, comparable to the observed value.

In the cool atmosphere, the electrons, which are required to couple the magnetic field to the gas, are supplied from the ionization of the heavy elements with low FIPs. As a result, in cases with low metallicity, sufficient electrons are not provided, causing severe damping of Alfv\'enic waves.
Therefore, the mass-loss rate positively depends on metallicity in the range of $Z\le 0.1Z_{\odot}$. In the metal rich side, $Z>0.1Z_{\odot}$, the $\dot{M}$ is saturated because the radiative cooling is enhanced as the density in the atmosphere increases.

In the atmosphere of these 
late-type giants, there simultaneously exist 
cool gas with $T\approx$ a few $10^3$ K and warm gas with $T \sim 10^4-10^5$ K. In addition, in $\alpha$ Boo, magnetized hot bubbles with $T\gtrsim 10^6$ K are also formed in a time-dependent manner. Although these hot bubbles are present only 
for limited durations in the single flux tube that we are modeling ($\approx 1/6$ of the simulation time in the fiducial case), 
we expect that in the realistic 3D atmosphere they ubiquitously appear in a stochastic manner, which is consistent with the detection of X-rays in $\alpha$ Boo \citep{Ayres2003,Ayres2018,Schmitt2024}. 
In contrast, we do not find hot gas with coronal temperatures in $\alpha$ Tau with the magnetic field strength inferred from the observations, which is consistent with the fact that there is still no clear detection of X-rays at the moment \citep{Ayres2003}. However, the case with the stronger magnetic field exhibits transient magnetized hot bubbles. Additionally, we should note that, in order to quantitatively analyze X-ray emissions, it is necessary to consider closed magnetic loops, as the plasma confined in closed structures can be heated up more effectively to higher temperatures. 

The coexistence of multi-temperature gas triggers larger time variability in the mass-loss rate, which is typically about a few orders of magnitude in our simulations for a single flux tube. In reality, however, the mass-loss rate is defined as the integrated average of the mass flux from multiple open flux tubes covering the surface.
The time variation in the spatially averaged mass-loss rate would be 
reduced from that in the single flux tube. 

Our simulations can be directly applied to other late-type giants and supergiants that simultaneously possess cool MOLspheres and warm chromospheres. A straightforward application is to understand the stellar-mass dependence of such multi-temperature atmospheres of dust-free late-type (super)giants in a unified manner. For example, $\gamma$ Dra, which has almost the same effective temperature as $\alpha$ Tau but a higher mass, emits larger UV flux and drives 
{a faster wind} \citep{Robinson1998,Wood2016}, suggesting that the stellar mass is an essential parameter that governs the magnetic activity.
Extending our simulation to higher-mass, dust-free K supergiants is also important for understanding the magnetic effects on the chromosphere in a wide 
stellar-mass range \citep{Eaton1993,Wiedemann1994,Harper2013,Rau2018,Harper2022AJ,Nielsen2023}.

\begin{ack}
We thank Graham Harper for valuable comments and insightful suggestions to the original manuscript.
We are also grateful to the referee for constructive comments, which helped improve the manuscript.
This work is supported by Grants-in-Aid for Scientific Research from the MEXT/JSPS of Japan, 22H01263.
K.O. acknowledges the support of the Agencia Nacional de Investigaci\'on Cient\'ifica y Desarrollo (ANID) through the FONDECYT Regular grant 1240301.
\end{ack}


\bibliographystyle{aasjournal}

\end{document}